\documentclass[useAMS,usenatbib]{mn2e}

\title[The formation of molecular clouds in spiral galaxies]
{The formation of molecular clouds in spiral galaxies}
\author[C. L. Dobbs, I. A. Bonnell, J. E. Pringle]
{C. L. Dobbs$^1$\thanks{E-mail:
cld2@st-and.ac.uk}, 
I. A. Bonnell$^1$ \& J. E. Pringle$^2$\\
$^1$SUPA, 
School of Physics and Astronomy, University of St Andrews, 
North Haugh, St Andrews, Fife, KY16 9SS\\
$^2$Institute of Astronomy, Madingley Road, Cambridge, CB3 0HA}
\usepackage{amssymb}
\usepackage{amsmath}
\usepackage{psfig}

\begin{document}

\date{\today}

\pagerange{\pageref{firstpage}--\pageref{lastpage}} \pubyear{0000}

\maketitle

\label{firstpage}

\begin{abstract}
We present Smoothed Particle Hydrodynamics (SPH) simulations of molecular
cloud formation in spiral galaxies. These simulations model the response of a
non-self-gravitating gaseous disk to a galactic potential. 
The spiral shock induces high densities in the gas, and considerable structure
in the spiral arms, which we identify as molecular clouds. 
We regard the formation of these structures as due to the
dynamics of clumpy shocks, which perturb the flow of gas through
the spiral arms.  
In addition, the spiral shocks induce a large velocity dispersion in 
the spiral arms, comparable with the magnitude of the
velocity dispersion observed in molecular clouds. 
We estimate the formation of molecular hydrogen, by post-processing our
results and  
assuming the gas is isothermal. Provided the gas is cold ($T\le100$ K), 
the gas is compressed
sufficiently in the spiral shock for molecular hydrogen formation to occur in
the dense spiral arm clumps.
These molecular clouds are largely confined to the spiral arms, 
since most molecular gas is photodissociated to atomic hydrogen upon leaving 
the arms.
\end{abstract}


\section{Introduction}
Molecular clouds are the sites of star formation in spiral galaxies
\citep{Reddish1975,Shu1987,Blitz1999}. As such, 
they play an important role in the governing the properties of galaxies and 
their evolution. 
The properties of molecular clouds have been extensively 
studied in terms of their emission, kinematics and internal
structures (e.g. \citealt{Dame1986,Solomon1987,Brunt2003,Wilson2005}). 
Unfortunately they are still poorly understood in terms of their 
formation and evolution.
For example, it is unclear whether the molecular gas is formed \textit{in situ}
from 
atomic gas or whether molecular clouds are due to the coagulation of 
pre-existing molecular gas \citep{Blitz2004,Pringle2001}. Another important question is whether
molecular clouds are long lived or just very inefficient at forming stars
\citep{Elmegreen2000,Hart2001}. 
Inefficient star formation can be
potentially explained if the clouds are not globally self-gravitating 
\citep{Clark2004,Clark2005}.
In this case, locally bound regions can
explain the star formation while the vast majority of the gas escapes the 
cloud as it is unbound.

Molecular clouds are found to be primarily located in spiral arms of galaxies 
with an arm to interarm ratio varying from a few in the inner
regions of some galaxies \citep{Vogel1988,Adler1996,Reuter1996,Brouillet1998} 
to greater
than 20 in the outer regions of the Milky Way \citep{Heyer1998,Digel1996}.
Since spiral arms are generally defined by the presence of star formation, this
concentration of GMCs (giant molecular clouds) to the spiral arms in the Galaxy
is not surprising.
The total fraction of gas in molecular form also varies from 2 per cent in M33
\citep{Engargiola2003} to 75 per cent or more in M51 \citep{Scoville1983, 
Garcia1993}. However, the detection of CO may be somewhat underestimated due to
the uncertainty in the ratio of CO(1-0) luminosity to H$_2$ mass (the 'X
factor')
\citep{Kaufman1999}. 
Interestingly, the mass spectra of GMCs in galaxies are very similar with 
$dN \propto m^{-\gamma} dm$, $\gamma$ in the range of 1.5 to 1.8 
\citep{Solomon1987,Heyer1998}.
Typical molecular cloud masses in the Milky Way are 
100-1000~M$_{\odot}$ \citep{Heyer1998},
though GMCs up to 10$^7$~M$_{\odot}$ have been observed in the Milky Way
\citep{Solomon1987} and M51 \citep{Rand1990}.

There are several possibilities regarding the formation of giant molecular 
clouds (see \citealt{Elmegreen1990} and references therein). Where the
content of molecular gas is high, e.g. M51 or the
centres of galaxies, GMCs could form through the agglomeration of pre-existing
molecular clouds (e.g. \citealt{Scoville1979, Cowie1980, Tomisaka1984, 
Roberts1987}). 
In this scenario, the ISM may be predominantly molecular, but 
GMCs are only observed where the gas has coalesced into larger structures 
heated by star formation.
However this is unlikely to occur in the outskirts of galaxies or for example, M33,
where little molecular gas is thought to exist aside from GMCs
\citep{Engargiola2003}. Then other processes, most
commonly gravitational/magnetic 
instabilities \citep{Elmegreen1979, Elmegreen1982, Balbus1985, Balbus1988, Elmegreen1989, 
Elmegreen1991, Elmegreen1994} 
or compression from shocks \citep{Shu1972, Aannestad1973, Holl1979, 
Koyama2000, Bergin2004} must be required to convert atomic to molecular gas. 
Recently, \citet{Blitz2004} note that the fraction of molecular gas may depend 
on a specific factor, such as the gas pressure, whilst the formation mechanism for
GMCs is the same regardless of the molecular gas content.  

The concentration of GMCs to spiral arms can be explained in one way 
by the increased
local density \citep{Kennicutt1989}. Alternatively spiral density waves may take
a more important role and lead to the direct triggering of star formation
\citep{Roberts1969,Bonnell2006}. Several models 
\citep{Bergin2004,Holl1979,Koyama2000} propose formation of molecular hydrogen 
through shock compression of the ISM. Thus shocks from spiral density waves
could account for the preference of molecular gas in the spiral arms and
the significance of spiral arms as sites of star formation.  

The properties of molecular clouds include significant structure on all 
observed scales and supersonic bulk
motions that follow the relation $\sigma_v \propto R^{1/2}$
\citep{Myers1983,Dame1986,Brunt2003,Elmegreen2004}.  
These kinematics are generally
taken to be turbulence and thought to generate the internal structures. 
An alternative explanation
is that structure in the interstellar medium generates the internal velocity 
dispersion as it passes through a clumpy spiral shock \citep{Bonnell2006}.

\citet{Wada2002} have investigated self-gravitating galactic disks,
suggesting that self-gravity of the disk generates turbulence in the ISM.
The effect of
spiral shocks on the structure of non self-gravitating disks has also been
described \citep{Wada2004,Dobbs2006}. Recent simulations of gravitational
collapse in disk galaxies have investigated the dependence of the star formation
rate on the Toomre instability parameter and a Schmidt law 
dependence on the local surface density \citep{Li2005}. Molecular gas 
formation has generally been limited to smaller scale studies, e.g. cloud
collapse \citep{Monaghan1991} and colliding or turbulent 
flows \citep{Koyama2000,Audit2005}. 

In this paper, we look at the dynamics of a non-self-gravitating galactic disk
subject to a spiral density wave. We show the formation of clumpy molecular
cloud structures in the spiral arms. 
From \citet{Bergin2004} we take a simple equation for the conversion of
atomic to molecular gas which is applied to our galactic simulations.
We identify the location of molecular hydrogen in the galactic disk and
show the densities required for molecular gas formation.

\section{Calculations}
We use the 3D smoothed particle hydrodynamics (SPH) code based on the version by
Benz \citep{Benz1990}. The smoothing length is allowed to vary with space and
time, with the constraint that the typical number of neighbours for each particle 
is kept near $N_{neigh} \thicksim50$.  
Artificial viscosity is included with the standard parameters $\alpha=1$
and $\beta=2$ \citep{Monaghan1985,Monaghan1992}. 

\subsection{Flow through galactic potential}
The galactic potential includes components for the disk, dark matter halo and
the spiral density pattern. The spiral potential is provided solely by the
stellar background, which contains considerably more mass than present in the
gas. Thus we ignore the gravitational feedback from the gas to the stellar disk.
However, as the background potential is time dependent, the transfer of angular
momentum between the potential and the gas is possible (Section~3.6).

We represent the disk by a logarithmic potential 
\begin{equation}
\psi_{disk}(r,z)=\frac{1}{2} v_o^2 \log\bigg(r^2+R_c^2+\Big(\frac{z}{q}\Big)^2
\bigg)
\end{equation}
e.g. \citet{Binney} that provides a flat rotation curve with
$v_0=220$~km~s$^{-1}$. The core halo radius is $R_c=1$~kpc and $q=0.7$ is
a measure of the disk
scale height. A further component is the potential for the outer dark matter
halo, solved from the density distribution 
\begin{equation}
\rho_{halo}(r)=\frac{\rho_h}{1+(r/r_h)^2}
\end{equation}
where $\rho_h=1.37 \times 10^{-2}$~M$_{\odot}$~pc$^{-3}$ is the halo density and $r_h=7.8$~kpc
the halo radius
\citep{Cald1981}.
For the spiral density pattern, we use a potential from \citet{Cox2002}
\begin{equation}
\begin{split}
\psi_{sp}(r,\theta,t)&=-4\pi G H \rho_{0} \exp(-\frac{r-r_0}{R_s}) \sum_{n=1}^{3}
\frac{C_n}{K_n D_n} \cos(n\gamma) \\
\text{where} \quad
\gamma&=N\bigg[\theta-\Omega_p t-\frac{\ln(r/r_0)}{\tan(\alpha)}\bigg], \\
K_n&=\frac{nN}{r \sin(\alpha)}, \\
D_n&=\frac{1+K_n H +0.3(K_n H)^2}{1+0.3K_nH}, \\
C(1)&=8/3\pi, \quad C(2)=1/2, \quad C(3)=8/15\pi
\end{split}
\end{equation}
The number of arms is given by N and $r_0=8$~kpc, $R_s=7$~kpc and $H=0.18$~kpc  
are radial parameters. The pitch
angle is $\alpha$, the amplitude of the perturbation $\rho_0$ and the pattern
speed $\Omega_p$. 
We take N = 4 for a 4 armed potential, and a pitch angle of  $\alpha=15^o$, 
comparable with the Milky Way \citep{Vallee2005}. The 4 armed potential
also allows for more frequent passages of the gas through the spiral arms.
The pattern speed, $2 \times 10^{-8}$~rad~yr$^{-1}$ leads to a co-rotation 
radius of 11~kpc.
The amplitude of the potential is $\approx 200$~km$^2$~s$^{-2}$, with 
$\rho_0=1$~atom~cm$^{-3}$. 
This corresponds to a perturbation of 3 per cent to the disk potential,
and an increase of 20~km~s$^{-1}$ as gas falls from the top to the bottom of the
potential well. 
Overall the disk is in equilibrium, as the rotational velocities of gas in the
disk balances the centrifugal force from
the potential. The disk is in vertical equilibrium supported by an initial
velocity dispersion. The resulting scale height of the disk is maintained
throughout the simulation.

This paper only considers how hydrodynamic forces and galactic
potential influence the flow. In particular we do not include the effect of self
-gravity, magnetic
fields and feedback from star formation. The processes of heating and
cooling of the gas are simplified by taking an isothermal equation of
state. This approach is taken as our main objective is to assess the dynamics of
spiral shocks rather than comprehensively model the ISM. 

\subsection{Initial conditions}
We initially place gas particles within a region of radius
5~kpc$ < r < $10~kpc. The disk also has a scale height $z\approx 100$~pc.
We allocate positions and velocities determined from a 2D test
particle run. The particles in the test particle run are initially 
distributed uniformly with
circular velocities. They evolve for a couple of orbits subject to the
galactic potential, to give a spiral density pattern with particles settled into
their perturbed orbits.

In the SPH initial conditions, we give the particles velocities in the $z$
direction from a random Gaussian distribution of 2.5\% of the orbital speed.
We add the same magnitude velocity dispersion in the plane of the disk.
The total mass of the disk is nominally $5\times10^8$~M$_{\odot}$, but as 
self-gravity is not included, the total gas mass can be scaled to higher or 
lower masses.
The gas is distributed uniformly on large scales with an 
average surface density of $\Sigma
\approx 2$~M$_{\odot}$~pc$^{-2}$ and average density of 
$10^{-2}$~M$_{\odot}$~pc$^{-3}$. For the higher resolution run, most 
gas in the interarm 
regions has densities ranging from $10^{-3}$~M$_{\odot}$~pc$^{-3}$ to 
$10^{-2}$~M$_{\odot}$~pc$^{-3}$ whilst the initial peak density in a spiral arm is 
$\approx 0.5$~M$_{\odot}$~pc$^{-3}$ (Figure~1).
The nominal surface density ($\approx 2$~M$_{\odot}$~pc$^{-2}$) is lower than
the gas surface density of 5~M$_{\odot}$~pc$^{-2}$ suggested by
\citep{Wolfire2003} for a radius of 8.5 kpc, although we mainly focus on 
modelling just the cold phase of the 
atomic gas.
In Section~3.4 we consider the formation of molecular gas with 
different disk masses and therefore surface densities.  

The number of particles 
is either $10^6$ or $4\times10^6$. The higher resolution work was
carried out on UKAFF (UK Astrophysical Fluids Facility) whilst the $10^6$
particle simulations were run on local computing facilities.
We ran simulations 
for between 200 and 300 Myr, where the orbital periods at 5 and 10 kpc are 150 
Myr and 300 Myr respectively. The gas passes through multiple spiral shocks in
this time and reaches a steady state.
No boundary conditions are applied to our calculations. 
With $4\times10^6$ particles, the disk is fairly well resolved in the spiral 
arms, where the ratio of the smoothing length ($h$) to the disk scale height 
($H$) is approximately 0.4 (the disk scale height calculated as the height of
the disk which contains 2/3 of the local disk mass). However the ratio $h/H$ 
increases up to $\approx 1$ in the interarm regions.   

All calculations are isothermal with temperatures of 10, 50, $10^2$, $10^3$, or 
$10^4$~K. 
Most previous simulations and analysis of spiral galaxies 
(e.g. \citealt{Wada2004,Kim2002,Dwark1996}) assume a sound speed of 
approximately 
10~km~s$^{-1}$ corresponding to the hotter ($10^4$~K) component of the ISM.
However, for GMCs to form, gas entering the spiral arms must be cold atomic
clouds \citep{Elmegreen2002} or pre-existing molecular gas \citep{Pringle2001}.
Since our aim of this work is to investigate molecular cloud
formation in shocks, lower ISM temperatures are adopted for most of these
calculations. We do not include self gravity, but estimate the Toomre
parameter for the disk to be $Q \thicksim 2.7$, and so the disk is 
stable against gravitational collapse. 
In any case, the gas will reach a spiral arm before gravitational 
collapse can occur e.g. \citep{Bonnell2006}. We refer to \citet{Li2005b} for 
a discussion of the stability of disks for different $Q$.
 
\section{Disc dynamics and cloud formation}
\begin{figure}
\centerline{\psfig{file=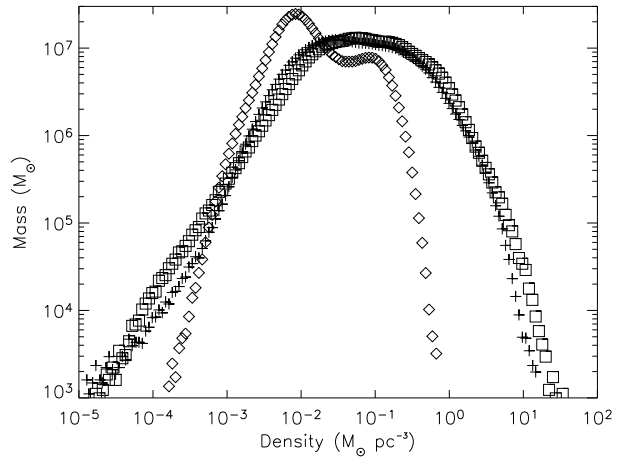,height=2.6in}}
\caption{The distribution of mass as a function of gas density at the beginning
(diamonds), after 100~Myr (crosses) and at the end (280~Myr) (squares) of the 
50~K simulation. The nominal density for molecular clouds is
2.5~M$_{\odot}$~pc$^{-3}$.}
\end{figure}
\begin{figure}
\centerline{\psfig{file=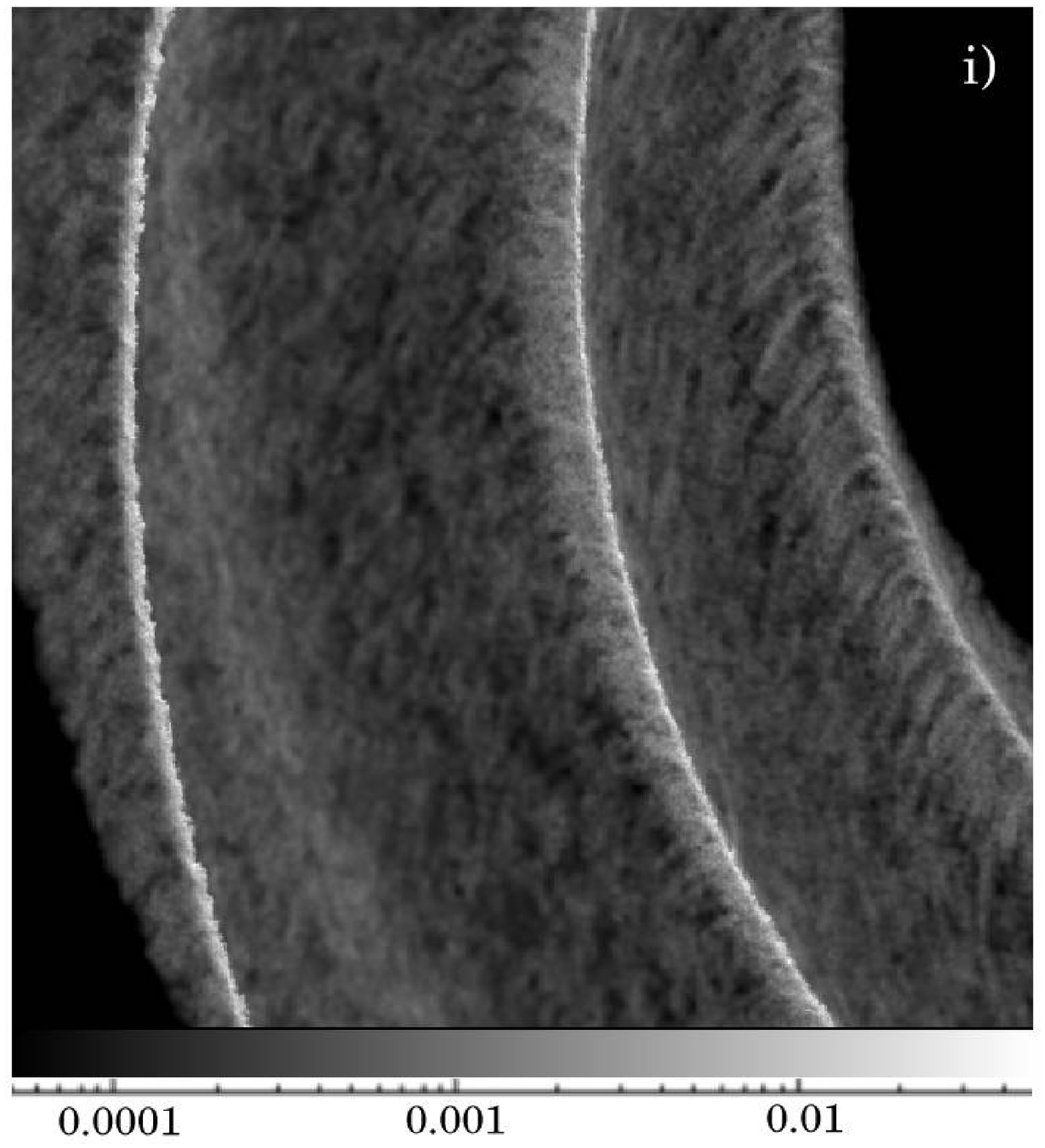,height=2.8in}}
\centerline{\psfig{file=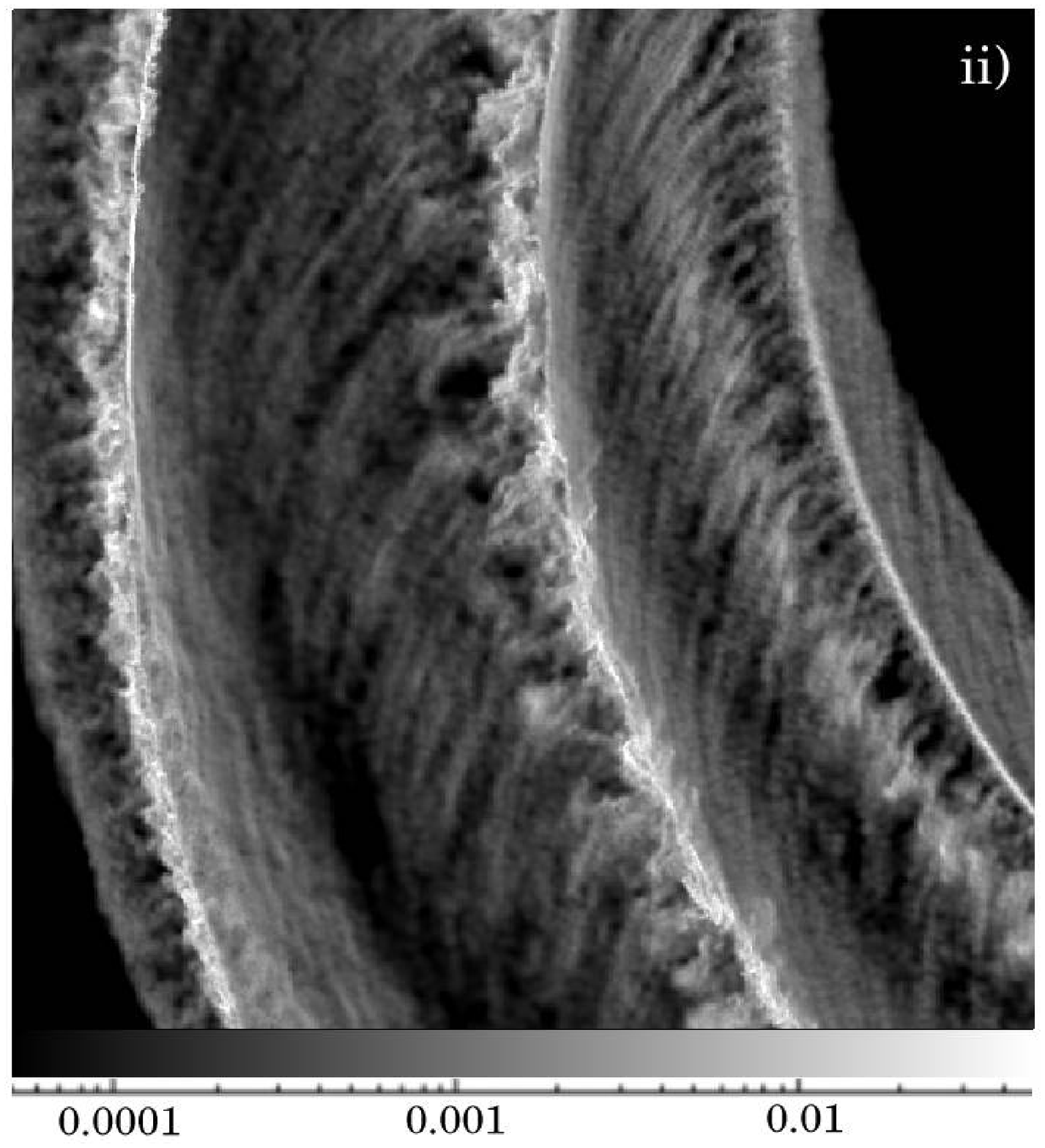,height=2.8in}}
\centerline{\psfig{file=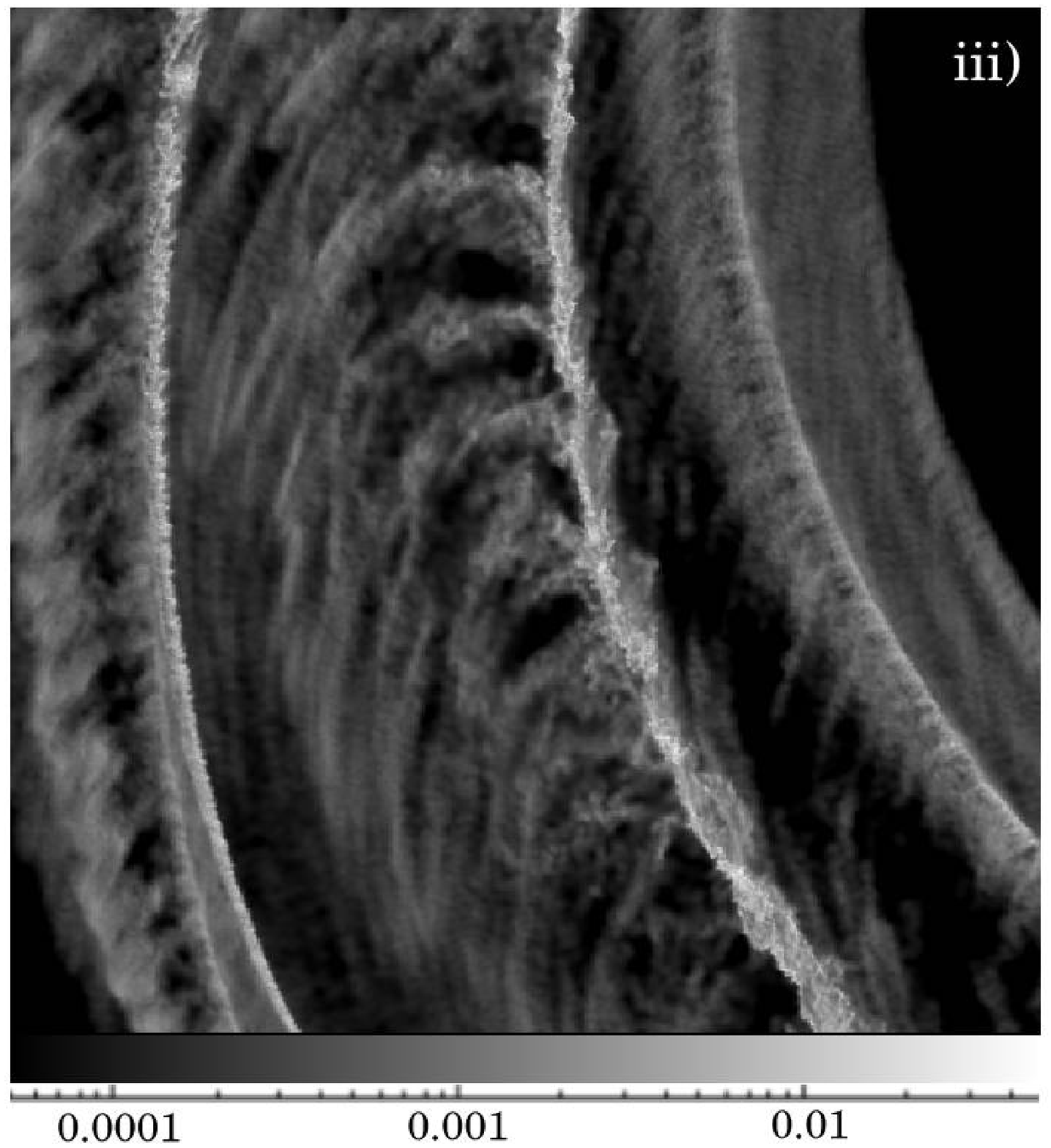,height=2.8in}}
\caption{Column density plots (g~cm$^{-2}$) showing a 5 kpc by 5 kpc section of 
the disk (with Cartesian coordinates -9 kpc $< x <$ -4 kpc and -5 kpc $< y <$
0 kpc. The xy-coordinate grid is centred on the midpoint of the disk and remains
fixed with time. 
Gas is flowing clockwise across the disk, i.e. from bottom to top.
The time corresponding to each plot is i)~50~Myr, ii)~100~Myr and iii)~150~Myr.
This simulation used 4 million particles with a temperature
of 50~K.}
\end{figure}
We first describe briefly the overall structure for the different simulations 
performed. Section~3.1 then examines the formation of structure in the spiral
arms, before we discuss the velocity dispersion across the disk and the
distribution of molecular gas.  

Initially, the disk is approximately
uniform with a smooth spiral perturbation. 
The widths of the spiral arms are even along the length of each arm, and the
densities across the arms are fairly uniform.
As the shock develops,   
the density increases in the spiral arms. 
In the simulations with cold gas,
substructure develops along the spiral arms (Figure~2). The gas in the arms becomes clumpy,
and the arm width fluctuates. Densities of gas in the arms varies over
approximately two orders of magnitude. The spiral arms become more ragged, and
their width increases. Meanwhile dense clumps leaving the arms are sheared,
leading to the formation of spurs perpendicular to the arms 
\citep{Dobbs2006}.    
The distribution of gas densities reaches an approximate equilibrium  
after $\approx 100$~Myr (Figure~1) with gas densities in the spiral arms
reaching $> 10$ M$_{\odot}$ pc$^{-3}$.
In contrast, the higher temperature 
simulations show broader, less dense spiral arms (see \citet{Dobbs2006}). 
The spiral shocks are much
weaker for the higher temperature gas, since the change in velocity induced by
the spiral potential is more comparable with the sound speed of the gas. For the
$10^4$~K case, the sound speed is $\approx 10$~kms$^{-1}$ whilst the change in
velocity across the spiral potential is $\approx 20$~kms$^{-1}$ (Section~2.1). 
The 1000~K simulation shows
regular spaced more diffuse clumps along the spiral arms, but the $10^4$~K run 
is relatively featureless, with no obvious structure in the spiral arms.
The difference in structure is due to the different strength of
the shock and the spacing of the clumps is determined by the spiral arm dynamics
(Section~3.1 and \citet{Dobbs2006}).

\subsection{Structure formation in the spiral arms}
We associate the densest regions of our simulation (Figure~2) 
with molecular clouds, 
as they attain the required densities and, 
as we shall see in Section 4, have appropriate physical
conditions and lifetimes that they should be primarily molecular gas.
The formation of dense molecular cloud structures can be understood as
being caused by the dynamics as interstellar gas passes through a spiral shock.
Initially, the pre-shock gas contains small-scale structure due to the particle
nature of SPH. The
spiral shocks amplify any pre-existing structures such that the clumpiness of
the gas increases with each passage through
a spiral arm. This amplification of clumpiness corresponds to changes in the
angular momentum phase space density of the gas. The pre-shock phase space
density is relatively smooth whilst the shocked gas displays much more structure
(Figure~3).
We provide an explanation of 
the generation of clumpy structure in terms of this change in the angular
momentum of the gas each time it passes through a spiral shock.
\begin{figure}
\centerline{\psfig{file=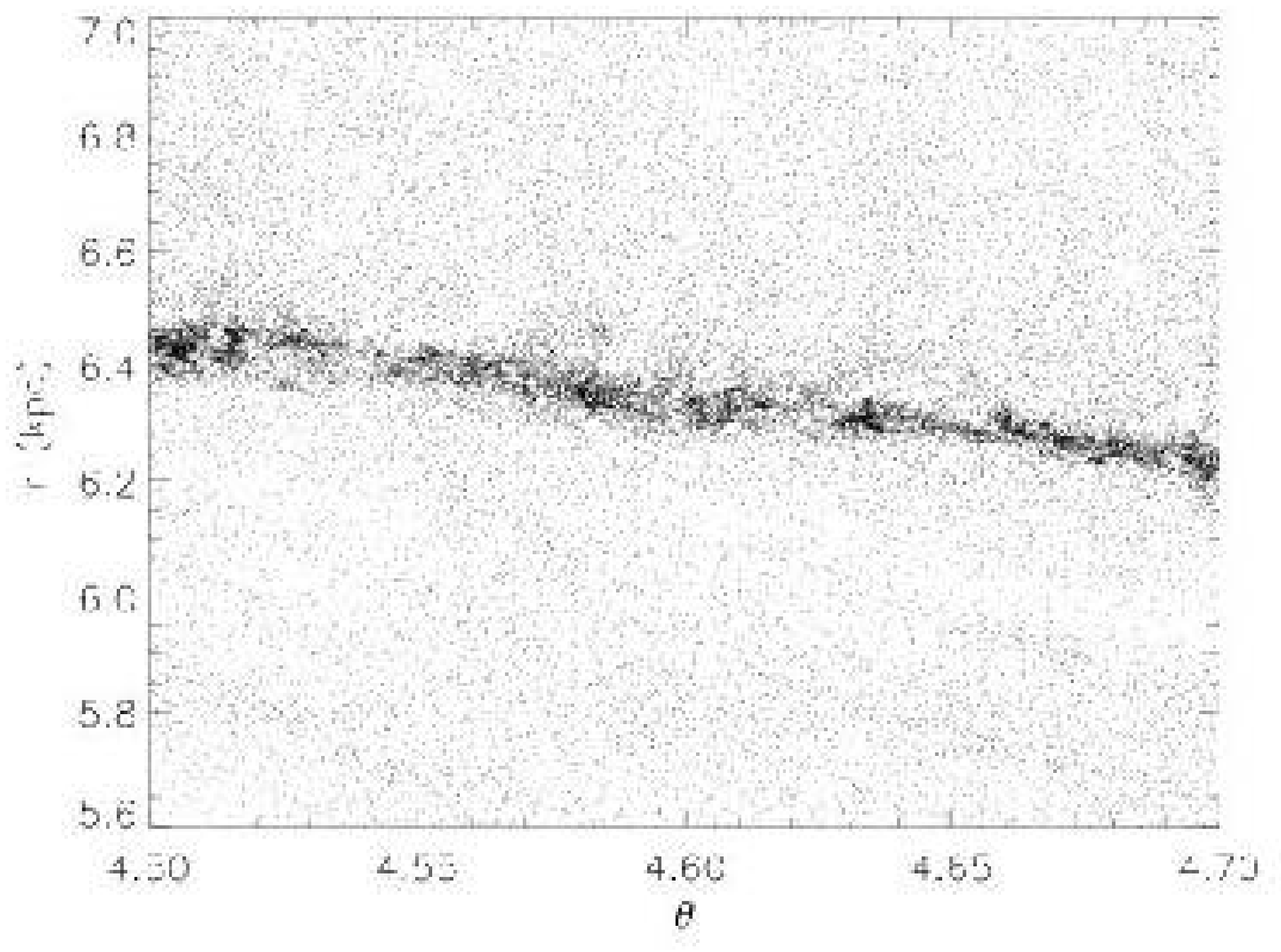,height=2.45in}}
\centerline{\psfig{file=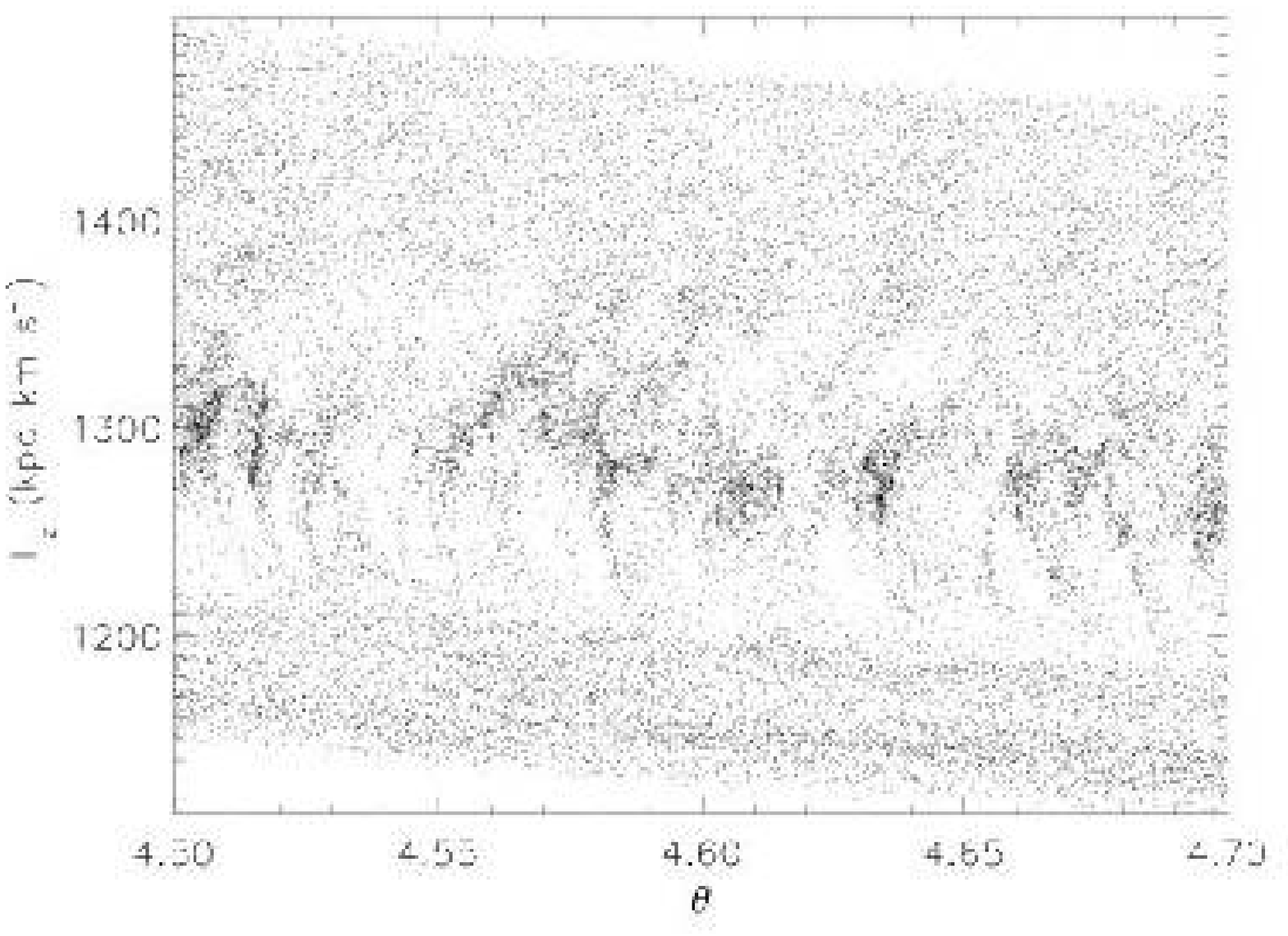,height=2.45in}} 
\caption{The top figure shows particles in a section of the disk in coordinates
($r,\theta$). $\theta$ is measured in radians clockwise round the disk, and 
thus the
section is taken from the top of the disk. The bottom figure shows the
corresponding $z$ component of the angular momentum plotted against $\theta$.  
The evolution of the structure in the angular momentum phase space is evident in
the bottom plot. The clumpy region shows the particles in the spiral shock.
Gas below this has yet to pass through the shock and has a relatively 
smooth distribution.
The post-shock gas lies above the clumpy region.
The simulation has evolved for 50~Myr.}
\end{figure}
 
As the arms are regions of strongly enhanced density, it is
evident that gas from a range of radii must interact and shock at any
particular point in a spiral arm. These shocks disturb the initially
smooth phase-space density of the gas particles by modifying the
particles' angular momenta. In Figure~4 we illustrate this by
plotting the evolution of $L_z$, the $z$-component of the angular
momentum, for a select number of particles as a function of
time. Initially the particles are at the same azimuth, but are at a
range of radii, with of course larger radii corresponding to larger
$L_z$. We see that particles with higher initial values of $L_z$, and
therefore at larger radii, enter the shock earlier. This is because
the spiral arms are trailing and the particles are inside the
co-rotation radius. As a particle enters such a shock it is typically
at the outer radial range of its epicyclic motion within the spiral
potential (see, for example, Figures 3 and 9 in \citet{Roberts1969}), and so
has lower angular momentum than the average for that radius. Thus when
it enters the shock, it mixes immediately with material which has
higher angular momentum. In Figure~4 the particles entry into the
shock is marked by a sudden jump in the value of $L_z$.  In Figure~5,
we show that the magnitude of the jump is correlated with the
post-shock density the particle encounters. Thus the stronger the
mixing, the bigger the jump. Each particle then travels along nearly
parallel to the shock for some distance, before it leaves the shock
\citep{Roberts1969}. As it moves along the shock, it stays roughly in phase
with the rotating spiral potential, and is subject to a torque from that
potential. Because these particles are inside co-rotation, the phasing
of the arm relative to the potential ensures that the torque is a
retarding one, and thus leads to a decrease in the value of
$L_z$. Once the particle leaves one spiral arm it then proceeds at
more or less constant $L_z$ until it meets the next one. As we would
expect for particles inside co-rotation, we can see from Figure~4
that the decrease in $L_z$, which occurs as the particle slowly leaves
the spiral arm, exceeds the increase it acquires on entry to the
arm. Thus the net effect of the spiral potential, coupled with
dissipation in the arms, is to produce a steady transfer of angular
momentum from the particles to the potential. This steady trend of
reducing the angular momentum of the particles is illustrated in
Figure~6.
 
We note that in the simulations with higher values for the temperature
of the ISM, the shocks are weaker, and the distribution in $L_z$ stays
much more uniform. This is because, after a weaker shock the particles
velocity is less affected and so the particle stays in phase with the
rotating potential, and so subject to a retarding torque, for a
smaller amount of time.
\begin{figure}
\centerline{\psfig{file=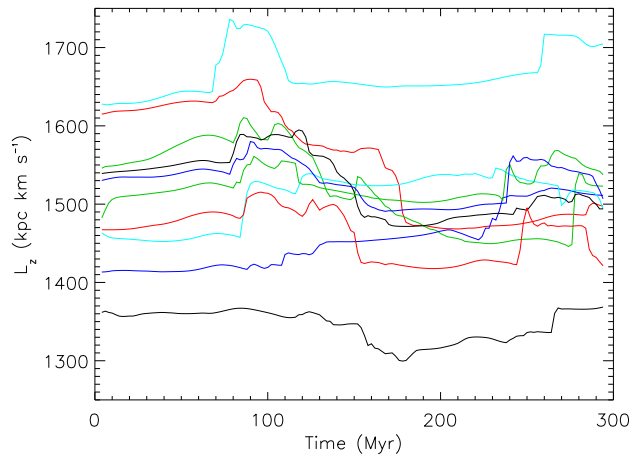,height=2.6in}}
\caption{The angular momentum ($z$-component) with time of 10 selected particles 
with 6.8~kpc $< r <$ 7.8~kpc. The particles are initially all at the same
azimuth, but at different radii.}
\end{figure}
\begin{figure}
\centerline{\psfig{file=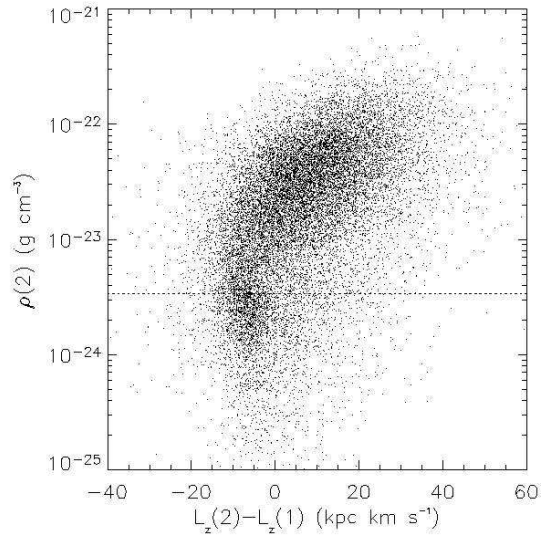,height=3.0in}}
\caption{The jump is $z$-component of angular momentum ($L_z(2) - L_z(1)$) is
plotted as a function of post-shock density ($\rho(2)$) for a number of
particles. The pre-shock densities of the particles (not shown) all
fall below the dotted line. There is a strong correlation between
density enhancement in the shock and the magnitude of the jump in $L_z$.}
\end{figure}
\begin{figure}
\centerline{
\psfig{file=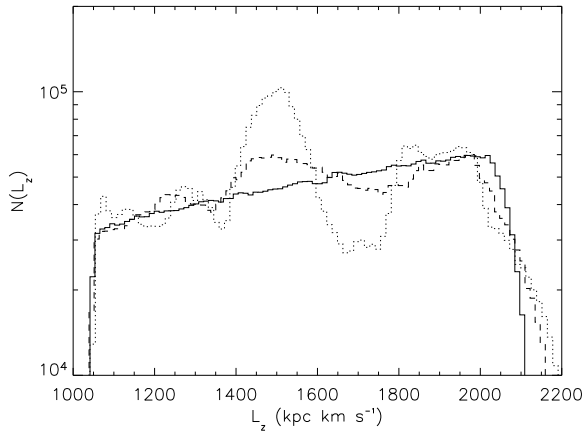,height=2.45in}}
\caption{The number of particles (N(L$_z$)) which have the value of the 
$z$ component of the angular momentum L$_z$. The total number of particles for the
whole disk is included at the beginning of the simulation (solid line), after
140~Myr (dashed line) and after 280~Myr (dotted line).}
\end{figure}
 
Although the pre-shock ISM is inhomogeneous, we find that this
clumpiness is amplified within the spiral arms. It is this
amplification which gives rise to the structures within the arm (Figure~2),
particularly in the molecular gas, as will be shown in Section~4.2. To see
this in more detail, Figure~3 (upper) shows the distribution of particles in a 
small sector of our simulation. Since $\theta$ is measured clockwise round the 
disk, particles are located in the top part of the disk, and are selected to 
cover a spiral arm. Approaching
the spiral arm from below can be seen apparent waves of
material. These are in fact the remnants of clumps formed in the
previous arm, by now strongly elongated by the differential shear in
the inter-arm region (See Figure~2). These inhomogeneities, which are
transmitted from the previous arm, provide input for new
inhomogeneities in the next arm. In Figure~3 (lower) we plot the value
of $L_z$ for each of the particles in Figure~3 (upper) as a function
of the $\theta$-coordinate. The region below the arm in the upper Figure~3
now spans a relatively small range in $L_z$, and the phase-space
density here is relatively smooth. Just above this is a region where
the phase-space density drops. This corresponds to the rapid jump in
$L_z$, discussed above, which occurs as particles enter the spiral
shock. Above this we see a patchy structure of phase-space density
enhancements. It is clear that the inhomogeneity which occurs here,
within the arm, is much greater than that seen in the pre-shock gas.
 
The physical picture we have of what is happening here is as
follows. We have seen that the gas, on entering the arm and being
shocked, moves along the arm for a while (Roberts, 1969). It
eventually leaves it to enter the inter-arm region, when its angular
momentum is too high to follow the spiral arm to smaller radii.
Imagine, then, as a simple model for the gas flowing along the arm, a
set of masses flowing in a line along a smooth inclined tube under
gravity. If the gas were all homogeneous, we would model this by
adding masses to the tube at equal velocities, at a rate which is
uniform in time and space. To mimic material leaving the arm, we
remove masses from the tube at positions which depend on their
velocity (for example, we could remove particles when their velocities
exceed some particular value $V_{\max}$). In this case, at any one
time the density of masses along the tube would remain smooth and
independent of time. However, to model an inhomogeneous pre-shock ISM,
we would add masses to the tube in a non-uniform manner in space, and
with a range of velocities. In this case, as the masses move along the
tube they would tend (assuming the encounters are dissipative) to
bunch up, like traffic along a single lane highway. This bunching in
velocity space can be seen in Figure~3 (lower). Since as before, we
would remove masses from the tube at positions which depend on their
velocities, it is clear that in this case, masses would not only tend
to form clumps within the tube, but would also tend to be removed from
the tube in clumps. As these clumps leave the spiral arms, they are
sheared out to form the spurs and feathering seen in Figure~2.

Using these ideas we can obtain a rough estimate of the linear scale
of the inhomogeneities we expect to find in the spiral arms. The
typical time a particle spends within a spiral arm can be estimated
from Figure~4 as being approximately $ t_{arm} \sim 6 \times 10^7$
years. The typical velocity spread of particles within a spiral arm
can be seen from Figure~7 to be approximately $v_\perp \sim 10$~km~s$^{-1}$.
Then the average scale-length of density structure generated by
the piling up process would be $L \sim t_{\rm arm} v_\perp \sim 600$~pc. 
This agrees roughly with the mean distance between the
high-density peaks, most easily viewed from the molecular gas density
(yellow) along the arm in Figure~9(iv).

\begin{figure}
\centerline{\psfig{file=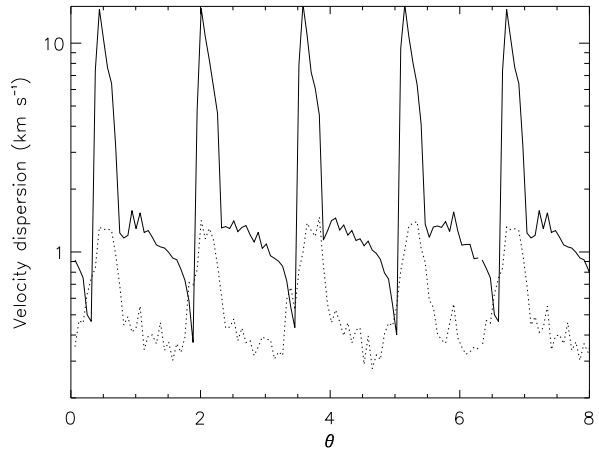,height=2.6in}}
\caption{The velocity dispersion plotted against azimuth ($\theta$, radians) 
for a ring whereby r=7.5~kpc and $\triangle$r=200~pc. The figure shows the
dispersion of the $v_{\perp}=\sqrt{(v_x^2+v_y^2)}$ (solid)
and the dispersion of $v_z$ (dotted). ($\theta$ is measured clockwise round the
disk, i.e. in the direction of gas flow). Insufficient resolution is likely to
produce damping in the interarm regions. The figure corresponds to a time of
100~Myr in the simulation.}
\end{figure}

\subsection{Velocity dispersion}
In addition to introducing structure in the phase space density of the gas,
spiral shocks can also induce large velocity dispersions in the gas
\citep{Bonnell2006}.
We determine the 2D velocity dispersion in the plane of the disk as a
function of azimuthal angle. This is calculated by taking a 200~pc width 
ring centred at $r=7.5$~kpc. We divide this ring into 100 segments, 
each of length $r\triangle \theta = 50$~pc and area $9\times10^4$~pc$^2$ and
calculate the velocity dispersion of the gas in each segment.
Figure~7 shows the variation of the velocity dispersion for the 50 K simulation.
The velocity dispersion is calculated
after 100~Myr, for the velocity component in the plane of the galaxy, 
$v_{\perp}=\sqrt{(v_x^2+v_y^2)}$, and the $z$ component 
of the velocity.
There is a large increase of 10~km~s$^{-1}$ in 
the dispersion of $v_{\perp}$ where the spiral arms occur.
In comparison a velocity dispersion of 7-10~km~s$^{-1}$ is observed for
molecular clouds of sizes 20-50~pc \citep{Solomon1987,Dame1986}.
\citet{Bonnell2006} find a velocity dispersion up to 10~km~s$^{-1}$ in
simulations of gas passing through a spiral shock. This velocity 
dispersion is due to
random chaotic internal motions induced in clumpy gas subject to a shock.     
The velocity dispersion quickly decreases upon leaving
the arms. A further decrease in the velocity dispersion occurs in the interarm
regions. The ratio in the smoothing
length compared to the disk scale height ($h/H$) approaches 1 in the interarm
regions. This potentially leads to damping of velocities due to insufficient
resolution, particularly in the $v_z$ velocity component.
 
\section{Molecular gas}
We have seen above that spiral shocks produce highly structured and dense gas
that can be associated with molecular clouds in spiral arms.  In addition, we
would like to know whether such a formation mechanism is able to account for the
formation of molecular gas from initially atomic gas or
if such gas would need to be primarily molecular before entering the shock. In 
order to
do this, we perform a rough estimate of the formation (and destruction) of 
molecular gas as the gas enters and leaves the spiral arms. We post-process the 
simulation by using the local gas properties determined
by the isothermal simulations and calculate the time dependent formation rate 
of H$_2$ on dust grains as well as the
dissociation rates by UV radiation and cosmic rays.
At each time step from the simulation, the formation and destruction rate is 
calculated for each gas parcel (particle)
and added to its previous molecular fraction. In this way, we can estimate the 
spatial and temporal evolution
of the molecular gas throughout the simulation.

Since the
gas in our models is isothermal, the formation of molecular hydrogen is not
treated self-consistently. In particular there is no heating or cooling, as
described in previous smaller scale analysis 
\citep{Bergin2004,Audit2005,Koyama2000}. 
As a first attempt to analyze the overall distribution of molecular hydrogen,
and the densities required for H$_2$ formation, we post-process our results to
find the time-dependent molecular density of gas in the disk. This is a
simplistic approach but identifies the most dense gas in our simulations where
molecular hydrogen forms. The reader should bear in mind
however that heating from
feedback and supernovae explosions will have some effect on these results.

We use a simple algorithm for the evolution of molecular hydrogen 
\citep{Bergin2004}
determined for molecular cloud formation behind shock waves.
The rate of change of H$_2$ density is given by
\begin{equation}
\frac{dn(H_2)}{dt}=R_{gr}(T)n_p n(H)-[\zeta_{cr}+\zeta_{diss}(N(H_2),A_V)]n(H_2)
\end{equation}
where $n$ is the number density (of atomic or
molecular hydrogen), $N$ is the column density (of atomic or
molecular hydrogen) and T the temperature. 
The total number density is $n_p$, where $n_p=n(H)+2n(H_2)$. Molecular hydrogen
is produced through condensation onto grains, determined by the first term on
the right of eqn(4). $R_{gr}$ is the formation rate on grains, which is proportional
to T$^{0.5}$. The efficiency for the formation rate on
grains is $S=0.3$ , although recent results suggest the efficiency is higher 
than this for temperatures of less than 90 K \citep{Cazaux2004}. 
Destruction of H$_2$ occurs through cosmic ray ionization and
photo-dissociation, represented by the second term on the right of eqn(4). The
cosmic ionization rate, $\zeta_{cr}$ is assumed constant
($\zeta_{cr}=6\times10^{-18}$~s$^{-1}$).   

Following \citet{Bergin2004}, we adopt the approximation of \citet{Draine1996} 
for the H$_2$ dissociation rate:
\begin{equation}
\begin{split}
\zeta_{diss}(N(H_2))&=\zeta_{diss} f_{shield}(N(H_2))e^{-\tau_{d,1000}}\zeta_{diss}(0) \\
f_{shield}(N(H_2))&=\frac{0.965}{(1+x/3)^2}+\frac{0.035}{(1+x)^{0.5}} \\
&\times \exp[-8.5\times10^{-4}(1+x)^{0.5}] \\
\text{where} \quad x&=N(H_2)/5\times 10^{14}\text{cm}^{-2}
\end{split}
\end{equation}
The function $f_{shield}$ approximates the 
rate of photodissociation summed over all energy levels. Self shielding is 
included through the dependence on column density and    
$\tau_{d,1000}$ is the dust extinction at 1000 \AA. The constant
$\zeta_{diss}(0)=4.17\times10^{-11}$~s$^{-1}$ is a measure of the strength of 
the UV field.  This
formulation assumes that the background UV spectrum is the same as
that of a B0 star, and that the radiation flux is constant
everywhere. The strength of the flux is measured by the magnitude of
$\zeta_{diss}(0)$. We note that this
is a very approximate measure of the the photodissociation rate, which should
ideally vary depending on the local level of star formation.   
Furthermore, we only produce an estimate of the
column density $N(H_2)$ in our calculations by multiplying the local density by the 
scale height of the disk.

We divide Eqn(4) by $n_p$ and calculate the change in the molecular gas 
fraction, $d(n(H_2)/n_p)$, for each particle  
after each timestep (2~Myr). Assuming $n(H_2)=0$ everywhere initially, we add
the change in the molecular gas fraction to the ratio $n(H_2)/n_p$ determined
for the previous timestep. At each time frame, this ratio in the number densities
can be converted to a volume density of molecular hydrogen.
As shown in Figure~8, Eqn(4) mainly selects the most 
dense particles, although the molecular gas fraction is also time dependent. 
Since the formation of H$_2$ is not instantaneous,
the molecular gas fraction increases with the time gas spends 
in the denser spiral arms.
\begin{figure}
\centerline{\psfig{file=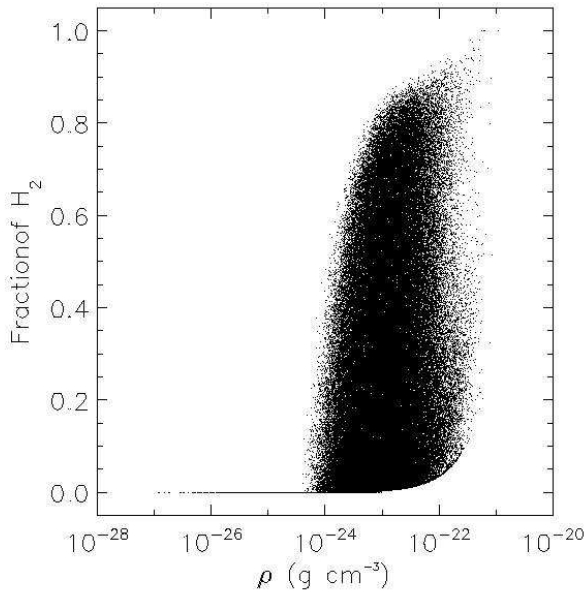,height=3.4in}}
\caption{The distribution of the molecular fraction of 
particles is shown versus
their density for the 50 K simulation, after 100 Myr. 1/10th of particles are 
plotted, i.e. $4\times10^5$.
There is a clear cut off at $n_p \thicksim 10^{-24}$ g cm$^{-3}$, below which
the gas is totally atomic. The large spread in the H$_2$ fraction for a given
density represents the time evolution of the H$_2$ fraction from when low
density gas first enters the spiral shock.}
\end{figure}

\subsection{Molecular gas formation at different temperatures} 
We find that H$_2$ formation in our simulations requires that the (isothermal) 
gas be cold, with temperatures of order 100 K or less. Of order 10\% of the gas
becomes molecular with these temperatures whereas when the gas is
1000 K the molecular fraction drops to around 1\% and is virtually zero at 
$10^4$ K.
We can thus conclude that the spiral shock formation of H$_2$ is likely when 
the ISM is cold.
The fraction of molecular gas was found to decrease with lower resolution.
For the 50~K highest resolution simulation, the total molecular gas in the 
disk reaches a maximum
of $\sim 13\%$. The fraction of molecular gas was found to decrease 
with lower resolution.
For 1 million particles, 
the amount of molecular
hydrogen formed was very similar in the 10, 50 and 100 K cases, reaching a 
maximum of $\sim 8\%$ of the total mass. 

The fact that the gas is assumed to be isothermal precludes a detailed study of the
temperature dependency on H$_2$ formation as in reality this would
depend on the details of the equation of state. For example, the efficiency
of grain formation, assumed here to be a constant S=0.3, can be as high as S=1 
for $T<20$ and as low as $S \sim 0.1$ for $T>100$ K \citep{Cazaux2004}.
What we can tell is that hot gas ($T\ge1000$ K)
does not form significant H$_2$ due to the shock dynamics.

When the gas is hot, the shock produced by the spiral potential is weaker, so
the density of gas in the spiral arms is reduced. The density in the $10^4$ K
simulation does not increase above $10^{-24}$ g cm$^{-3}$ which is too low for 
molecular
hydrogen to form (Figure~8). The density of gas 
in the spiral arms is
insufficient for self-shielding to occur and molecular gas is readily
photodissociated. Similar conclusions were reached by \citet{Bergin2004}, who
require pressures a few times higher than the average interstellar values to
allow effective self shielding. We do not include cooling,
although for strong shocks performed at $10^4$ K \citep{Koyama2000} 
only a thin layer of molecular hydrogen forms. Therefore at
higher temperatures, it is unlikely that spiral shocks could produce densities 
sufficient for significant molecular hydrogen formation.     

\subsection{High resolution simulation} 
In the remainder of the paper, we focus on the 50~K highest resolution
simulation.
We show the location of H$_2$ 
when the ISM temperature is 50~K in Figure~9. There are $4\times10^6$ particles
giving a mass resolution of 125~M$_{\odot}$ per particle. 
This figure is taken after
100~Myr, and shows the global distribution of molecular gas, and the small scale
structure of molecular clouds. 

Molecular hydrogen starts to form immediately in the spiral arms, the most dense
parts of the disk. H$_2$ is first situated in a narrow strip, where 
the shock from the spiral density wave occurs. The density of gas leaving the
arms is insufficient to prevent molecular gas quickly becoming
photodissociated, hence the molecular gas is solely confined to the shock in the
spiral arm.
This strip of molecular hydrogen is smooth and continuous along the spiral arm. 
However, as the density of the shock increases, denser structures become 
apparent in the arms. These break up into separate clumps of molecular gas,  
which we identify as 'molecular clouds'. These molecular clouds are 
produced simply by the kinematics of the gas flowing in the spiral potential 
as described in Section~3.1, as self gravity is not included. 
As the simulation progresses, dense structures leaving 
the spiral arms become sheared. Some regions of molecular gas 
leaving the spiral arms are then dense enough for self shielding to reduce the 
rate of photodissociation. These molecular clouds survive further into the 
interarm regions.
\begin{figure*}
\centering
\begin{tabular}{c c}
\psfig{file=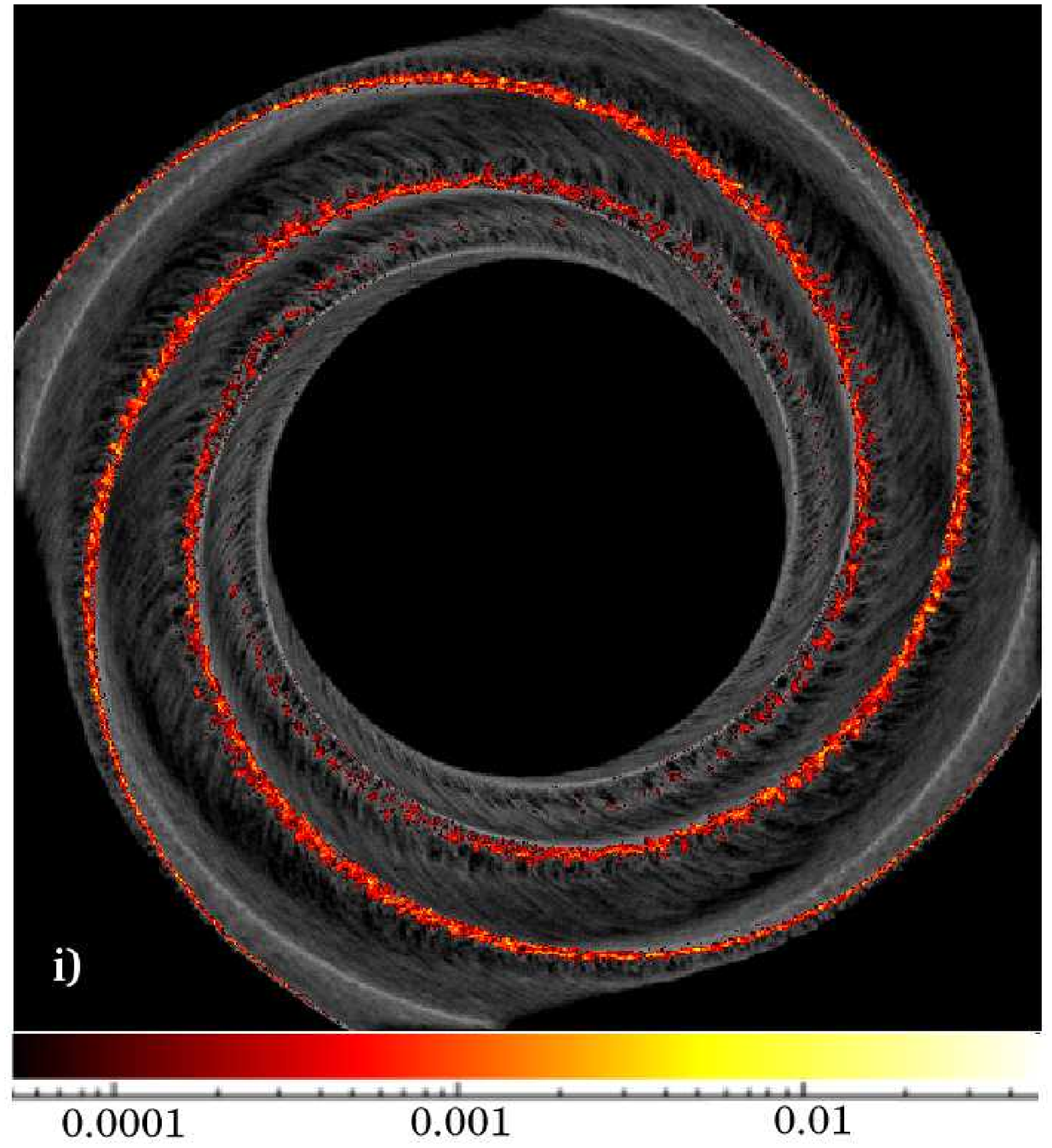,height=2.8in} & 
\psfig{file=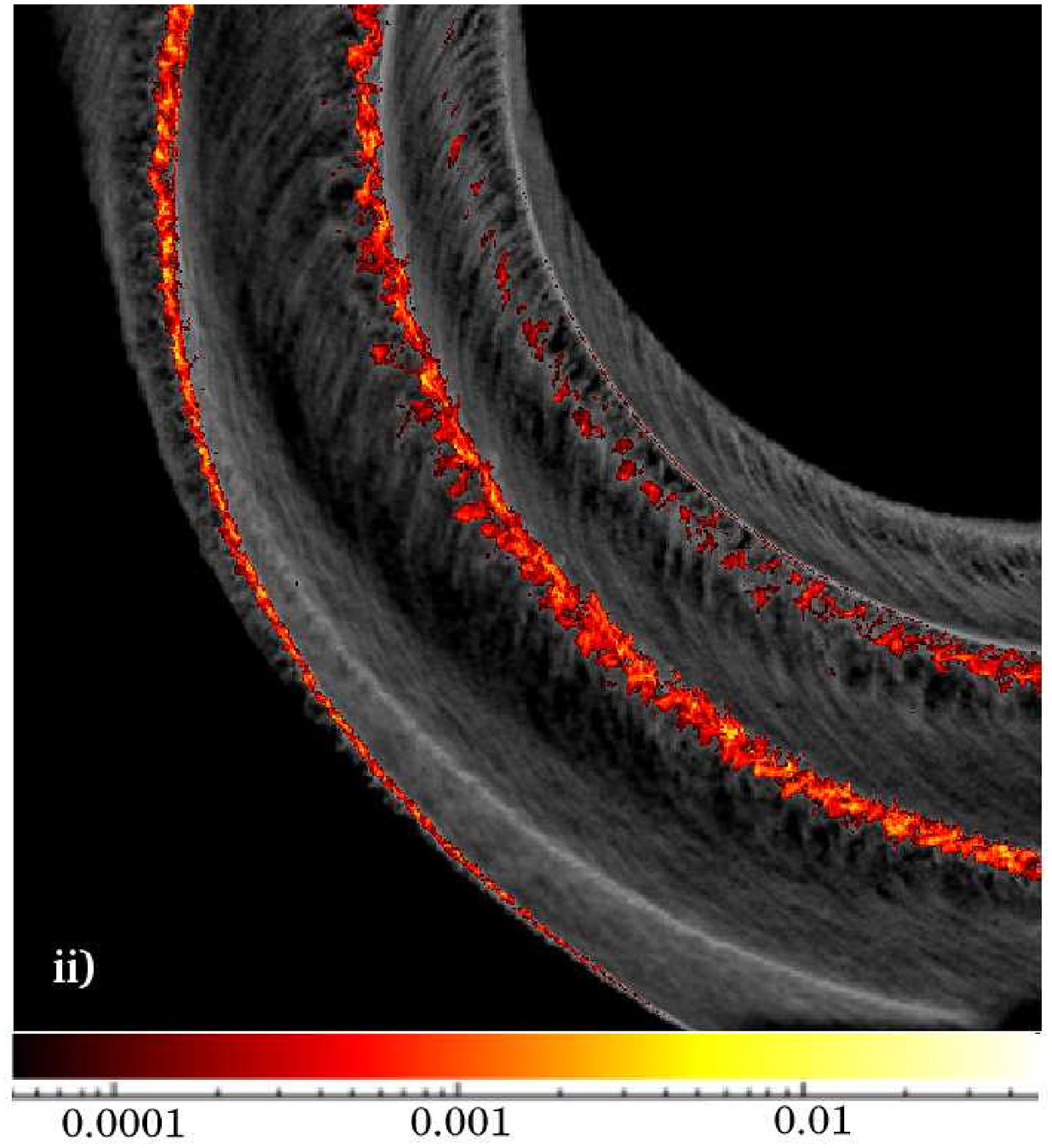,height=2.8in} \\
\psfig{file=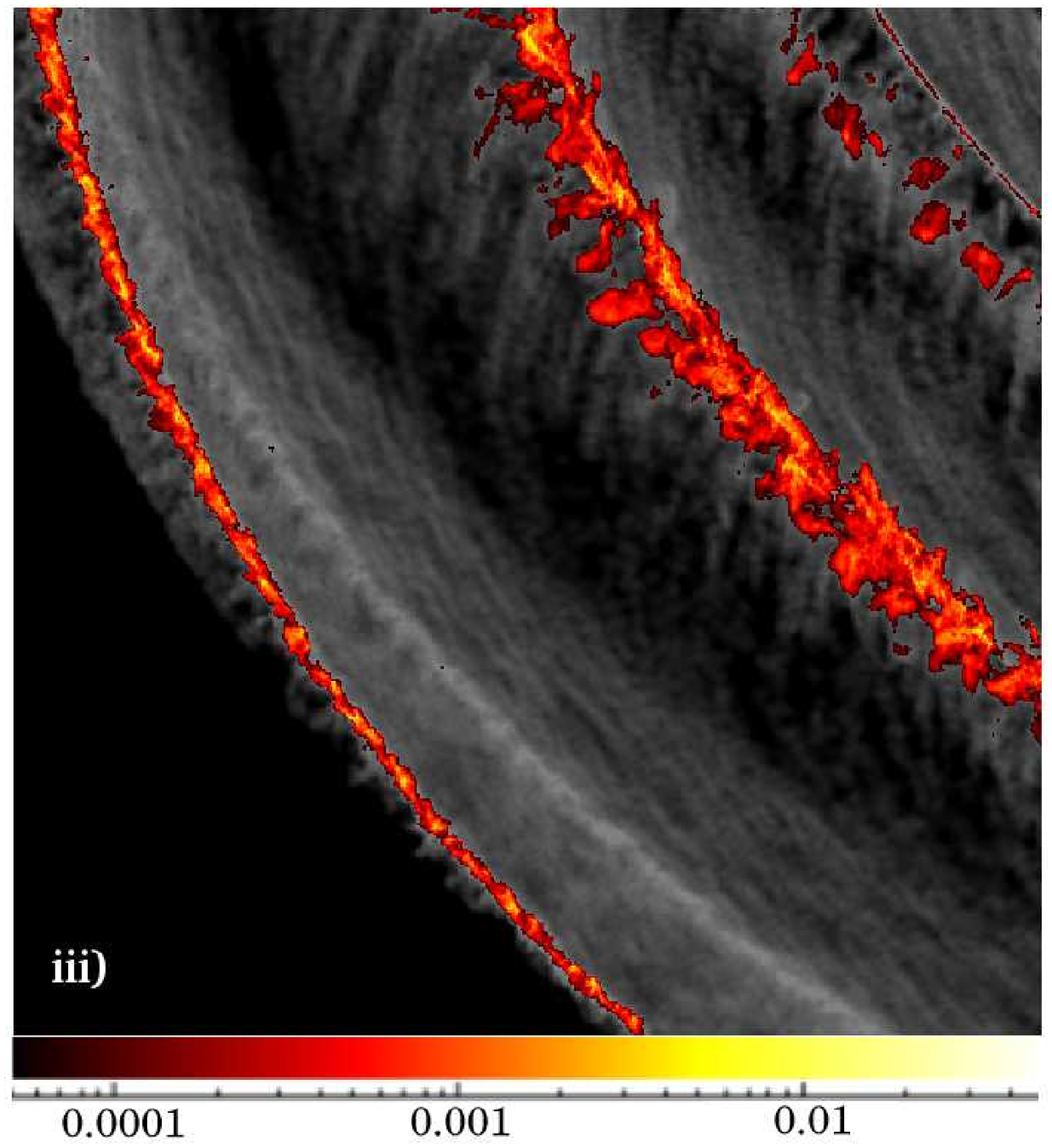,height=2.8in} & 
\psfig{file=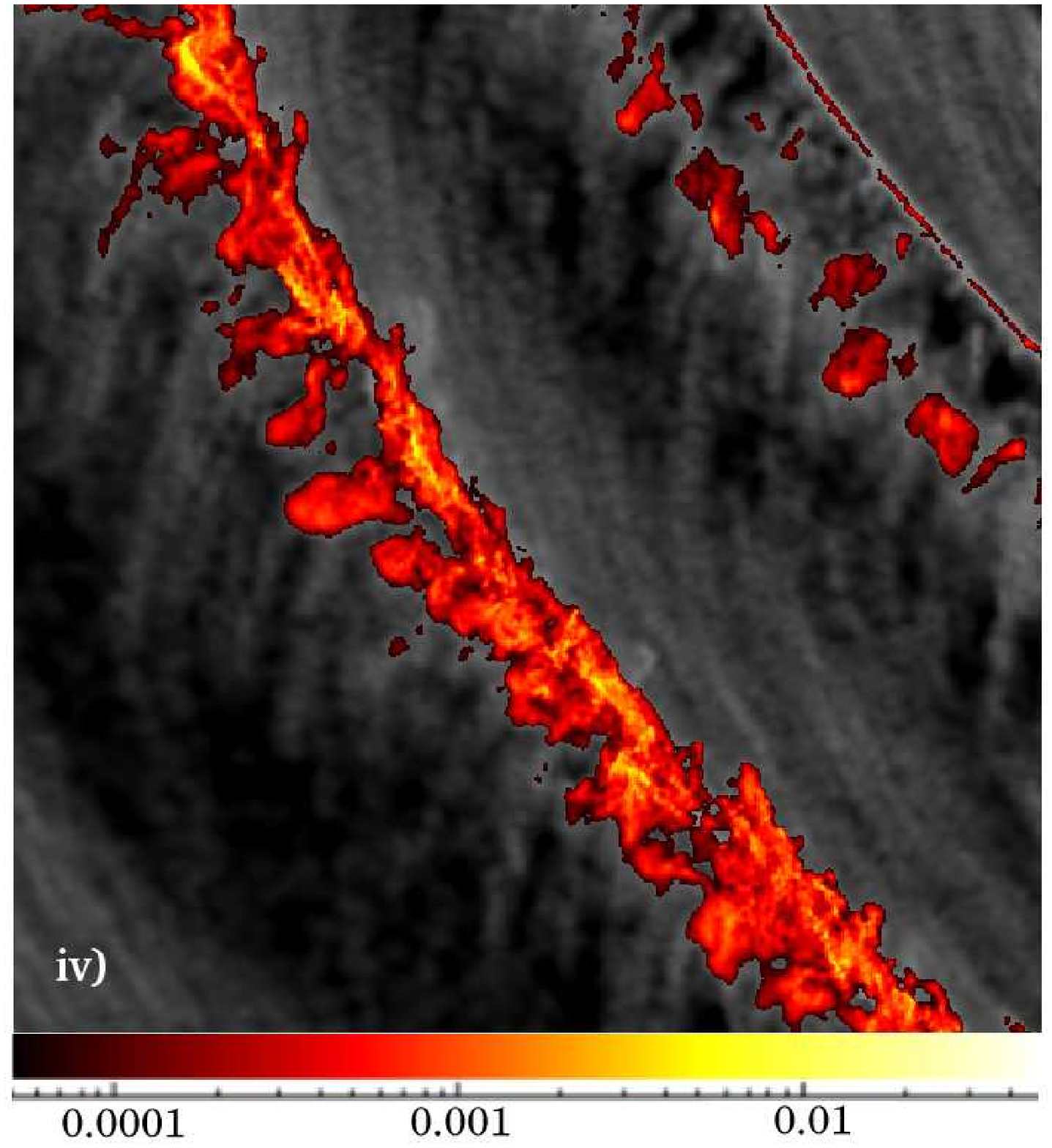,height=2.8in} 
\end{tabular}
\caption{Column density plots (g~cm$^{-2}$) 
showing density of molecular hydrogen (red) against
overall density (black and white).
Length-scales of plots are i) 20 kpc $\times$ 20 kpc, ii) 10 kpc $\times$ 10 kpc,
iii) 5 kpc $\times$ 5 kpc (-8.5 kpc $<x<$ -3.5 kpc, -8.5 kpc $<y<$ -3.5 kpc), 
iv) 3 kpc $\times$ 3 kpc (-6.5 kpc $<x<$ -3.5 kpc, -6.5 kpc $<y<$ -3.5 kpc).
The xy-coordinates assume a Cartesian grid centred on the midpoint of the disk 
which remains fixed with time.
Gas is flowing clockwise across the disk.}
\end{figure*}

\subsection{Molecular gas and spiral structure}
Figure~9 shows that molecular gas is largely confined to the denser spiral arms.
We plot the average molecular and total (molecular and atomic) gas densities
with azimuth in Figure~10, at 2 different times. As in Section~3.2, the densities
are calculated from a ring centred at $r=7.5$ kpc divided into 100 segments.
This plot only displays the mean molecular and atomic
densities - the peak densities are 10-20 times higher. 
The interarm densities of H$_2$ are $10^4$ times smaller than in
the spiral arms after 100 Myr, 
whilst the total densities are $\approx$ 50 times smaller. 
After 100~Myr, 10\% of the gas in the ring is molecular. 
Approximately 70\% of the total amount of gas is situated in the spiral arms, 
of which 20\% is molecular. Figure~10 suggests that the existence of interarm
molecular gas is unlikely, although we examine the effect of changing the disk
mass and photodissociation rate next.
\begin{figure}
\centerline{\psfig{file=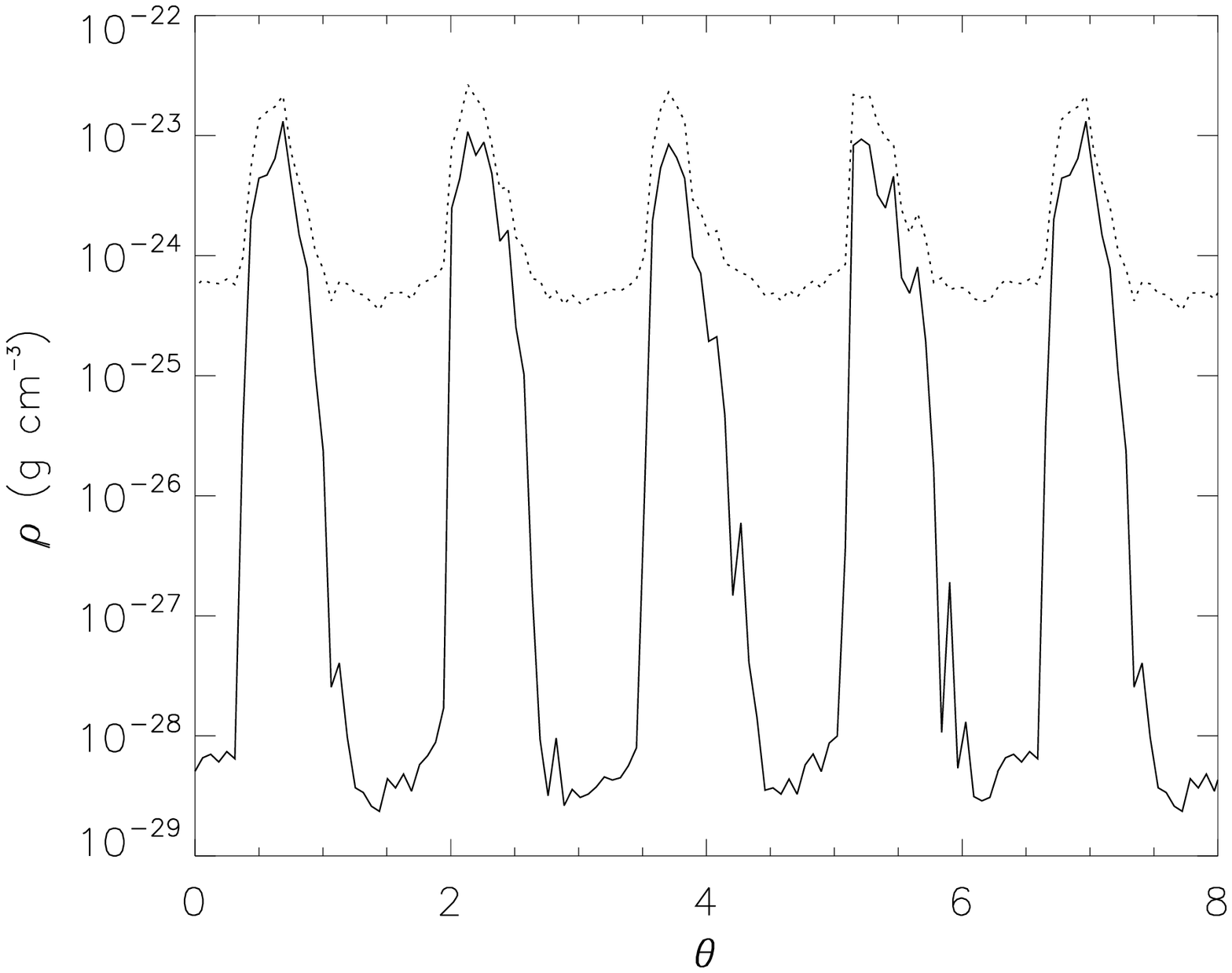,height=2.5in}}
\centerline{\psfig{file=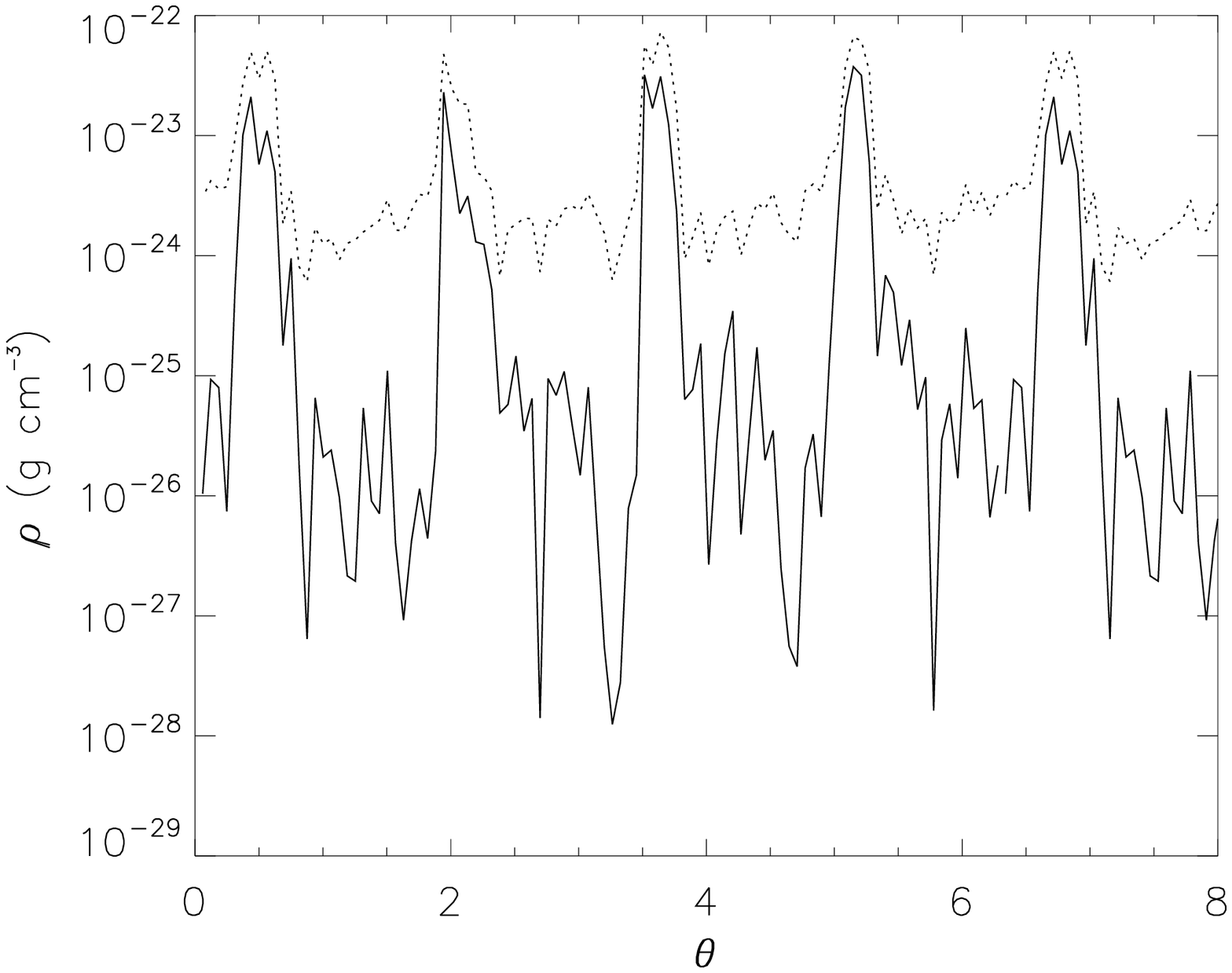,height=2.5in}}
\caption{Density of molecular gas (solid) and total density (dotted) are plotted
against azimuth ($\theta$), where $\theta$ (in radians) is the angle measured
in the direction of flow. The density plotted is the average over an annulus
centred at r=7.5~kpc with $\delta r$=200~pc.
The time of the simulation is 100~Myr (top) and 290~Myr (bottom).}
\end{figure}

\subsubsection{Dependence of H$_2$ formation on disk mass and photodissociation rate}
The mass of H$_2$ in eqn (1) is determined by the density, photodissociation
rate and temperature. The cosmic ray ionization rate,
$\zeta_{cr}=6\times10^{-18}$~s$^{-1}$, is found to have a minimal effect on the fraction of molecular
to atomic gas.
Since the H$_2$ mass is determined post process,
calculations can be repeated with a different disk mass (as there is no self
gravity) and/or different photodissociation rate.  
As expected, increasing the mass or decreasing the photodissociation rate
considerably increase the formation of H$_2$.
The mass of H$_2$ increases considerably to 
$1.4 \times 10^9$~M$_{\odot}$, 28\% of the total gas, when the disk mass 
increases by a
factor of 10. In comparing with the photodissociation rate, the same
approximation (eqn(4)) was applied, but increased or decreased by a constant
factor. When the
photodissociation rate decreases by a factor of 10, 
the molecular mass increases
to $1.25 \times 10^8$~M$_{\odot}$, 25\% of the total gas mass.
    
The disk mass has a substantial effect on the arm-interarm fractional abundance
of H$_2$. When the mass of the disk increases to $5\times10^9$~M$_{\odot}$,
nearly all
the gas in the spiral arms is molecular at the end of the simulation (Figure~11). 
Whilst the interarm ratio of the total density to H$_2$ density also increases, 
only about 1/10th of the interarm gas is molecular. 
The increase in both the arm and interarm regions 
is due to higher densities increasing the formation of H$_2$. 

The photodissociation rate has a more pronounced effect on the interarm ratios.
When the photodissociation rate is a tenth of that taken from \citet{Draine1996}, 
approximately a quarter of the interarm gas is molecular at the end of the
simulation (Figure~11). 
By contrast the amount of H$_2$ in the spiral arms changes much
less. This difference at a lower photodissociation rate 
arises when the gas is not fully dissociated between the spiral arms. Molecular
gas can survive for longer into the interarm regions leading to some molecular
structures between the arms. This is illustrated in Figure~12, which displays part
of the 50~K simulation where the molecular gas density is calculated with the
lower photodissociation rate. Clumps of molecular gas are shearing away from the
spiral arm. In addition, molecular material is arriving at the spiral arm which
has not fully dissociated from the previous spiral arm passage.

A more complete treatment of H$_2$ formation will give a more accurate idea of
the quantitative analysis of molecular hydrogen \citep{Glover2006a,Glover2006b}. 
Including heating may reduce the
amount of molecular gas, but we defer commenting on the effects of feedback, and
varying the photodissociation rate to depend on star formation to
future studies. These results indicate
nevertheless
that the molecular gas is strongly confined to the spiral arms. 
The possibility
that molecular hydrogen is transmitted between spiral arms cannot be ruled out
though, e.g. if the photodissociation rate is sufficiently low. 
The effects of increasing the disk mass or decreasing the level of
photodissociation are
likely to hold when including heating and cooling of the gas. 
\begin{figure}
\centerline{\psfig{file=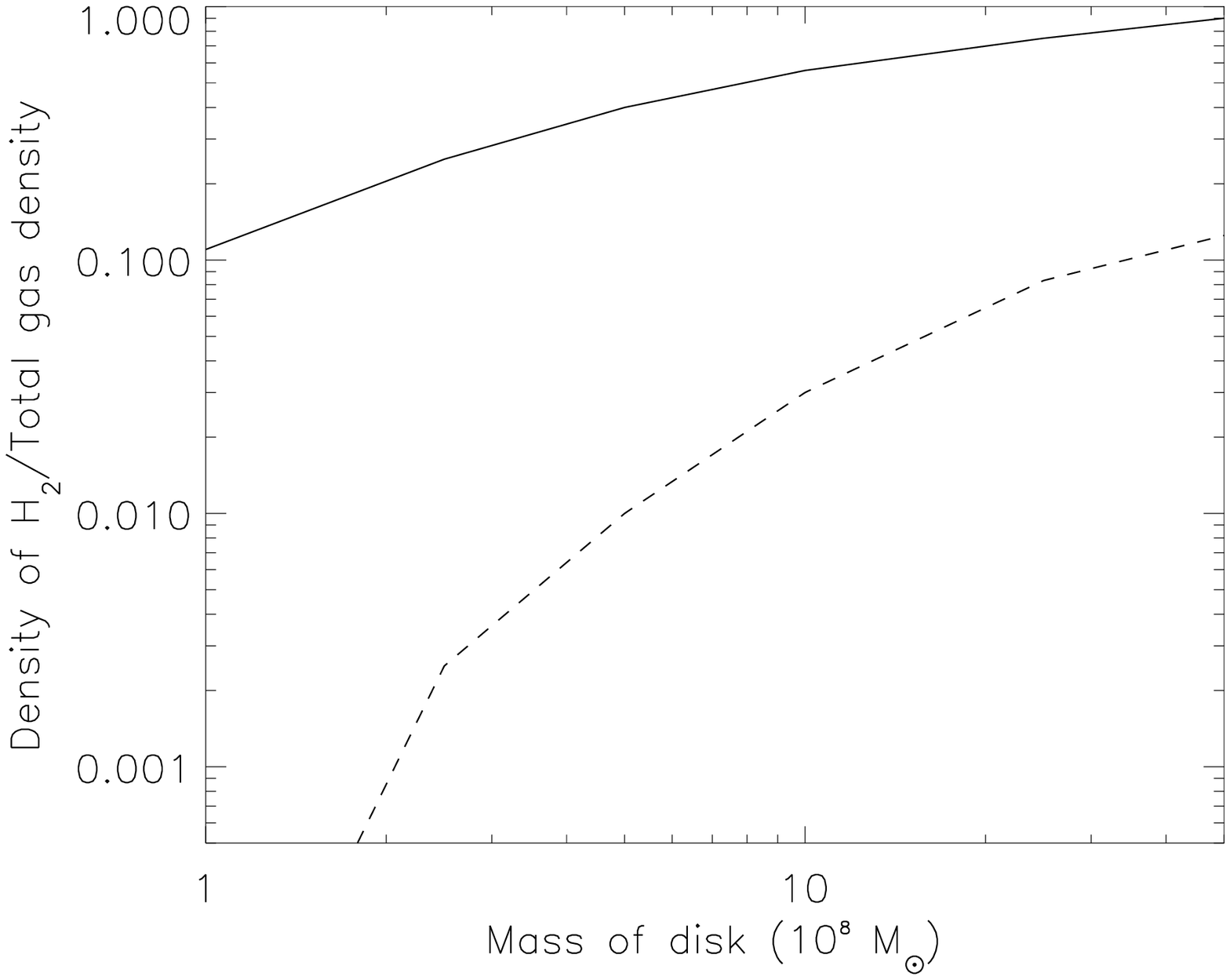,height=2.4in}}
\centerline{\psfig{file=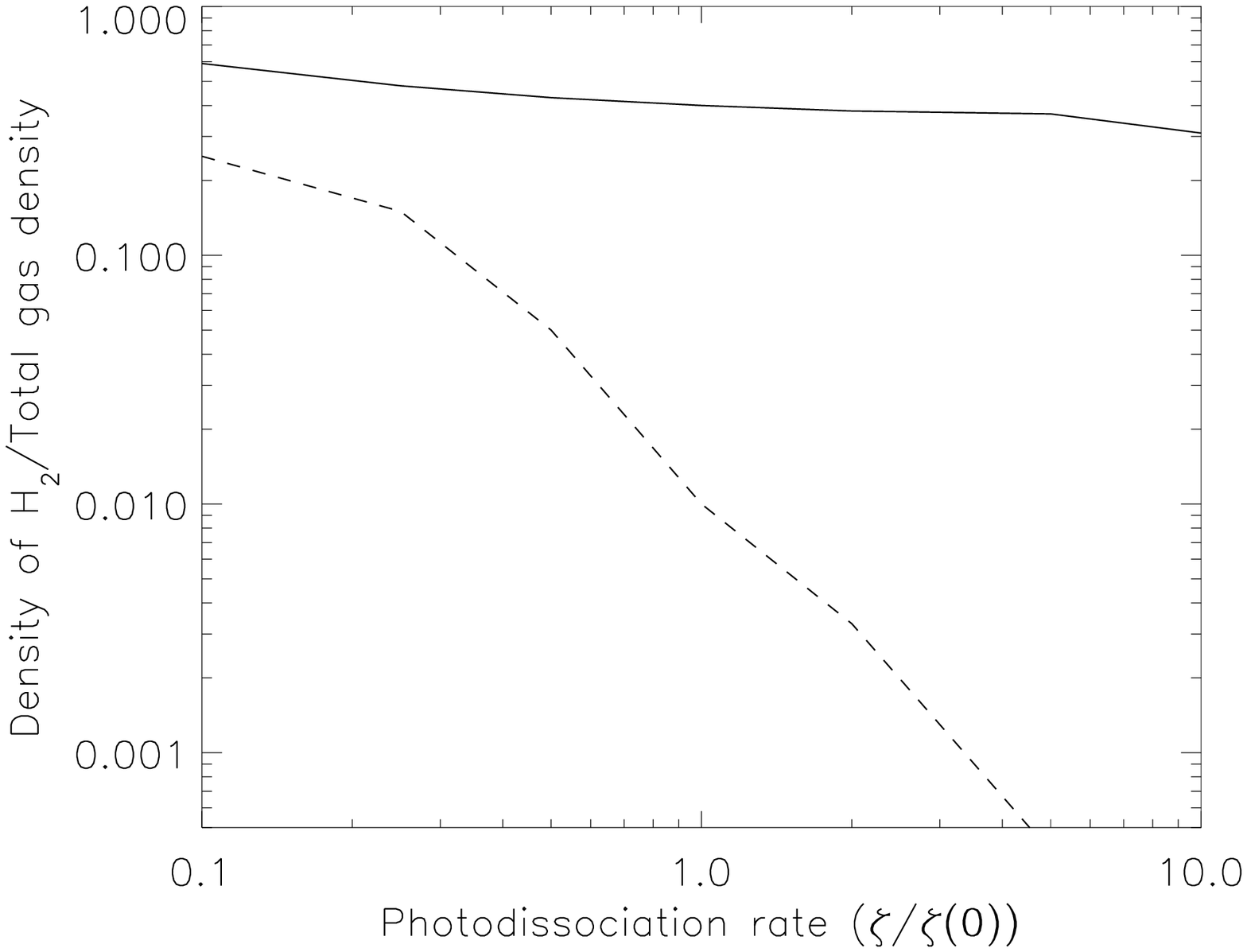,height=2.4in}}
\caption{Fractional abundance of H$_2$ in the arm (solid) and interarm regions 
(dashed) compared to the total disk mass (top) and photodissociation
rate (bottom). These figures are taken from the end of the simulation.}
\end{figure}

\begin{figure}
\centerline{\psfig{file=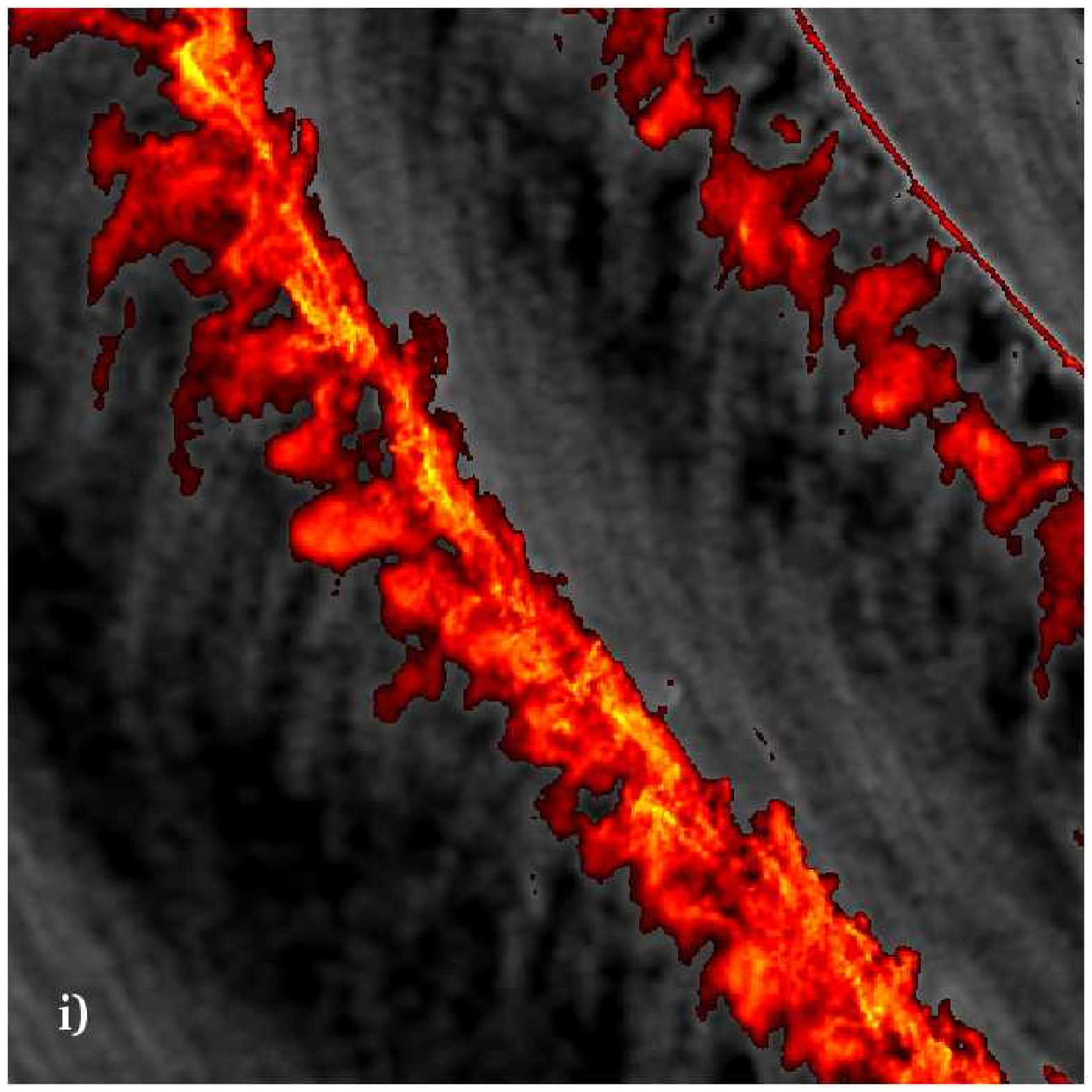,height=2.5in}}
\centerline{\psfig{file=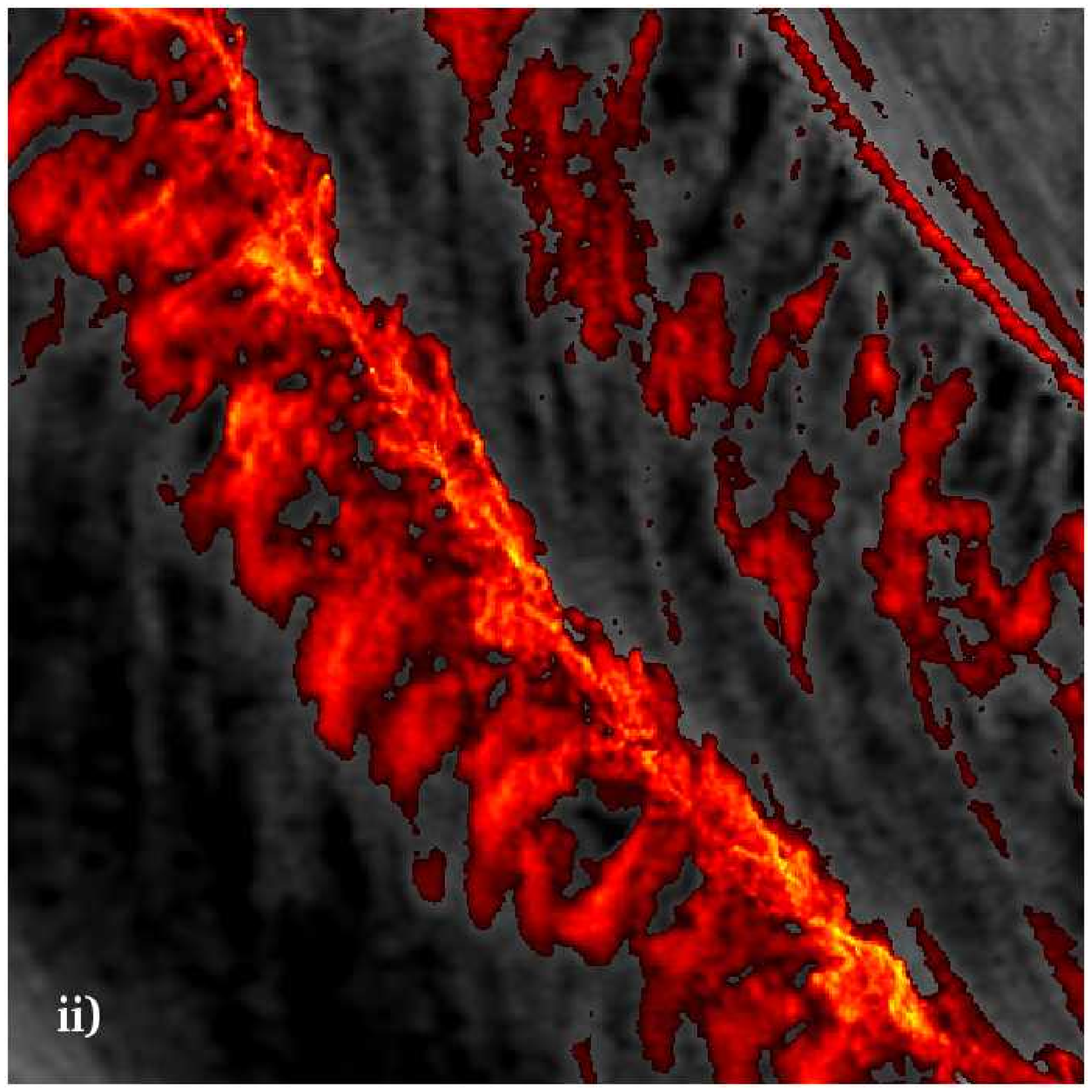,height=2.5in}}
\centerline{\psfig{file=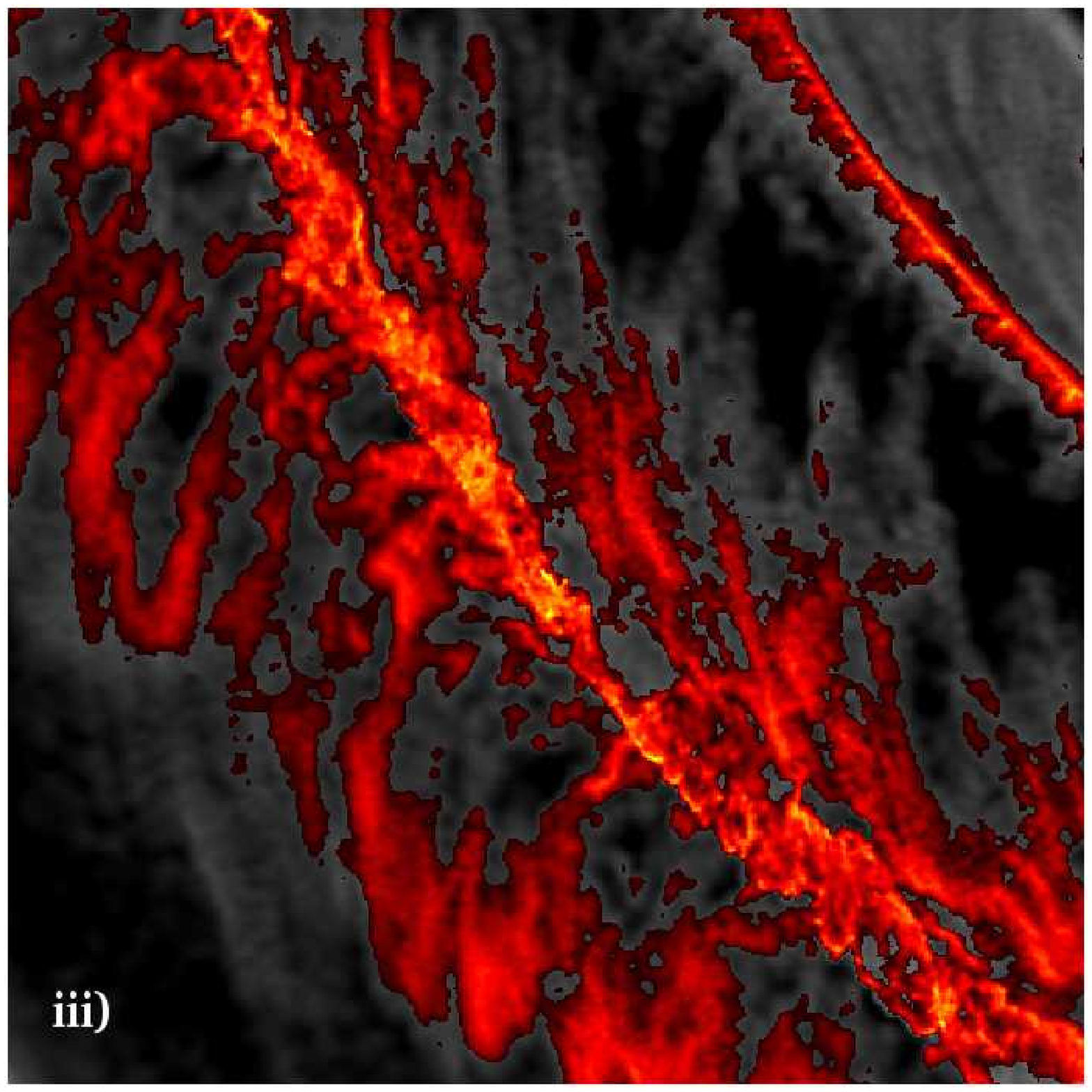,height=2.5in}}
\caption{A 3 kpc by 3 kpc region of the disk showing 
the transfer of molecular hydrogen between spiral arms.
The photodissociation rate is 1/10 that of
\citet{Draine1996} and as a consequence, there is more molecular hydrogen 
surviving between the spiral arms.
The column  density of molecular hydrogen (red) and total density 
(black and white) are displayed (same
scale as Figure~4). The times of each plot are i)100~Myr, ii)120~Myr and
iii)140~Myr. Plot i) is the same time, region and scale as Figure~4(iv), but using a lower
photodissociation rate. The increased density of H$_2$ along the arms leads to
molecular structures within the interarm regions which are absent from the
simulation in Figure~4.}
\end{figure} 

\section{Observational comparison of molecular cloud properties}
One of the objectives of these simulations is to see how well the spiral shocks
reproduce the observations of molecular clouds. We can directly compare our
results with molecular clouds in the Milky Way. We consider the distribution of
clouds from a local standard of rest comparable to the Sun, and a further LSR at
the edge of the disk. Several surveys of 
molecular clouds in the Milky Way have been undertaken
(e.g. \citealt{Dame1986,Dame2001,Lee2001}). We compare our data with 2 sets of 
results: i) the FCRAO CO survey of the outer galaxy \citep{Heyer2001}. 
This survey detected molecular clouds between the Galactic longitudes 102$^o$.49
and 141$^o$.54 which
are located predominantly in the local and Perseus arms; ii) a CO molecular
line survey \citep{Yang2002}, which detected cold infrared sources over the
northern plane of the Galaxy (longitudes between 0$^o$ and 260$^o$). 

Our calculations give the fraction of molecular gas for each particle in the
galactic disk. We use a simple algorithm to determine where molecular clouds are
located. We select a section of 1/8 of the disk as a sample, and 
divide this region into a 5~pc resolution grid. We calculate the total fraction
of molecular gas in each cell and retain cells
which contain over 10~M$_{\odot}$ of H$_2$. We classify all groups of 
adjacent cells satisfying this threshold as a molecular cloud. The properties
of each 'cloud', including mass, position and velocity are tabulated.
We determine the V$_{LSR}$ of each
molecular cloud - the locations of each LSR we 
have chosen are shown on Figure~13 compared to the overall distribution of
molecular clouds.    
\begin{figure}
\centerline{\psfig{file=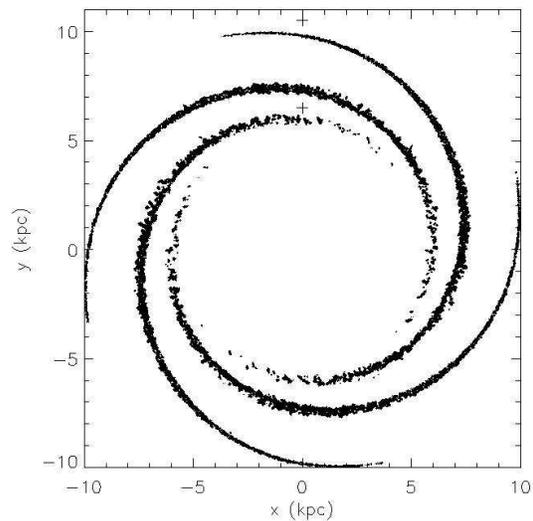,height=3.0in}}
\caption{The position of each local standard of rest (crosses) is compared to 
the overall distribution of molecular clouds. The LSR situated at (0,6.5)kpc is
comparable with the position of the Sun, in an interarm region in the mid-part 
of the disk. The second LSR (0,10.5)kpc produces a distribution of clouds
as viewed from the edge of the disk.}
\end{figure}
\begin{figure}
\centerline{\psfig{file=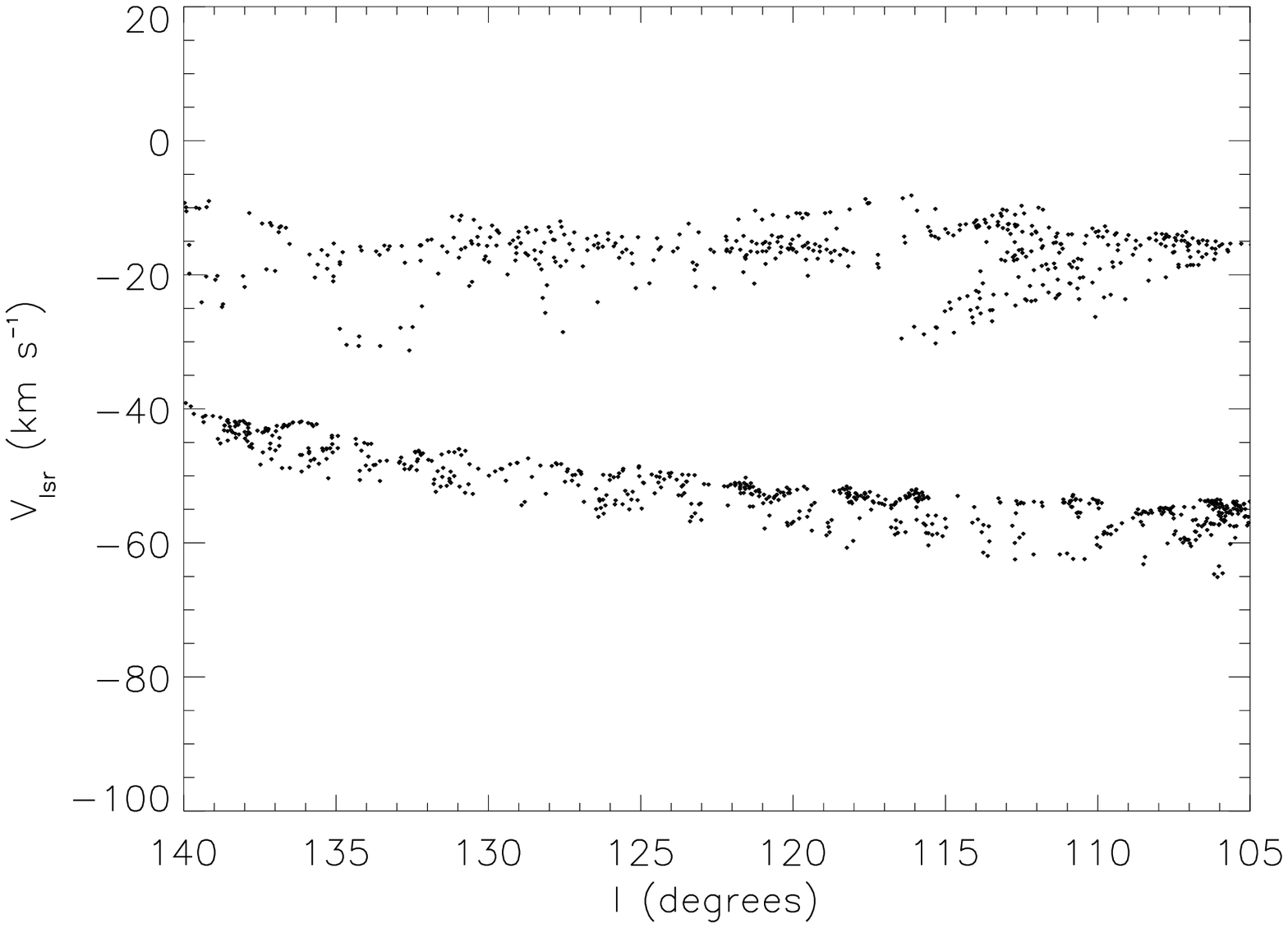,height=2.4in}}
\centerline{\psfig{file=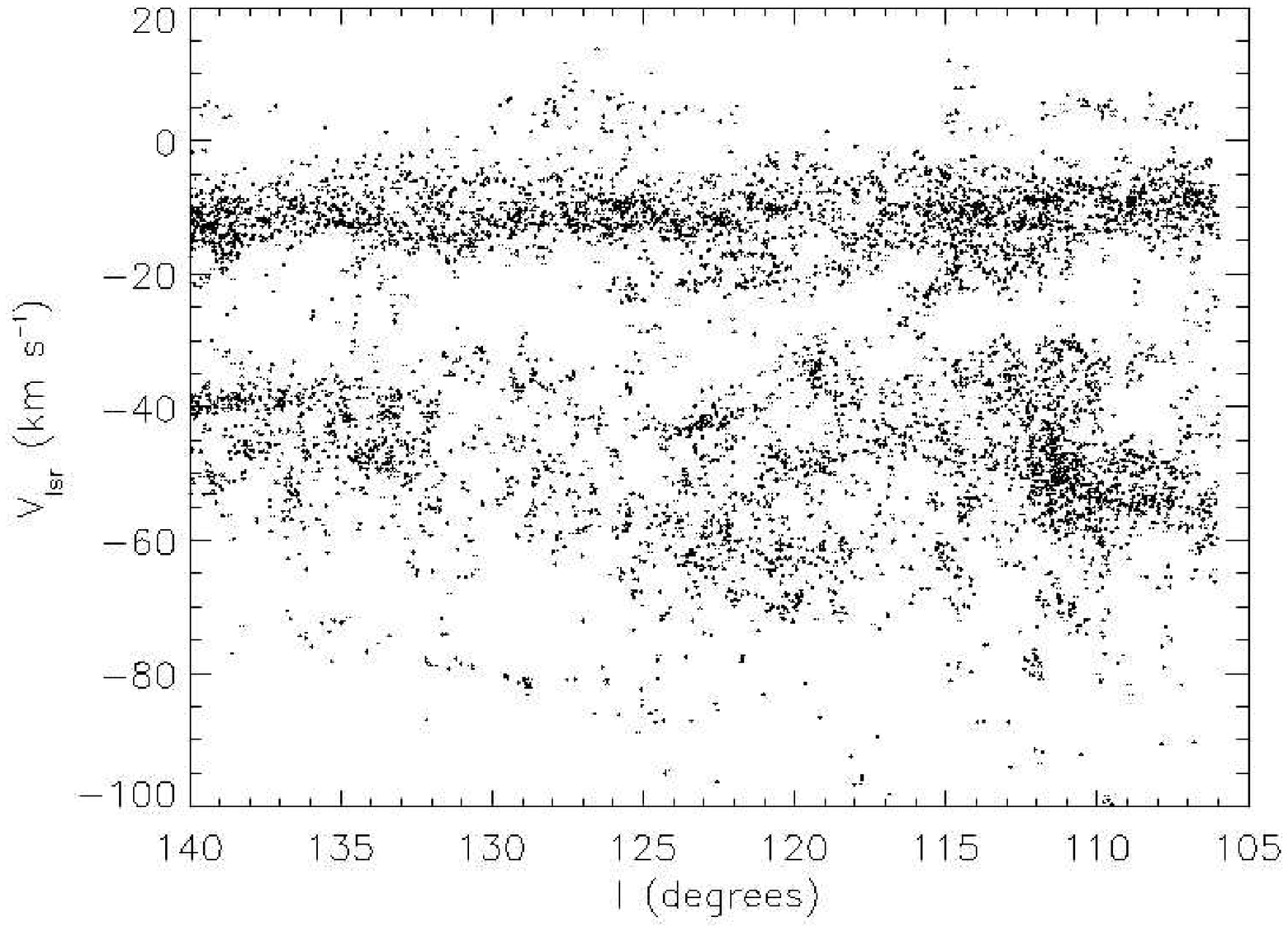,height=2.4in}}
\caption{V$_{lsr}$ plotted against galactic longitude for molecular clouds
from these simulations and the FRCAO CO survey. The top plot shows the simulated
data, with a LSR at coordinates (0,6.5)kpc. The lower plot uses data from the 
FCRAO CO survey \citep{FCRAO1998}.}
\end{figure}
\begin{figure}
\centerline{\psfig{file=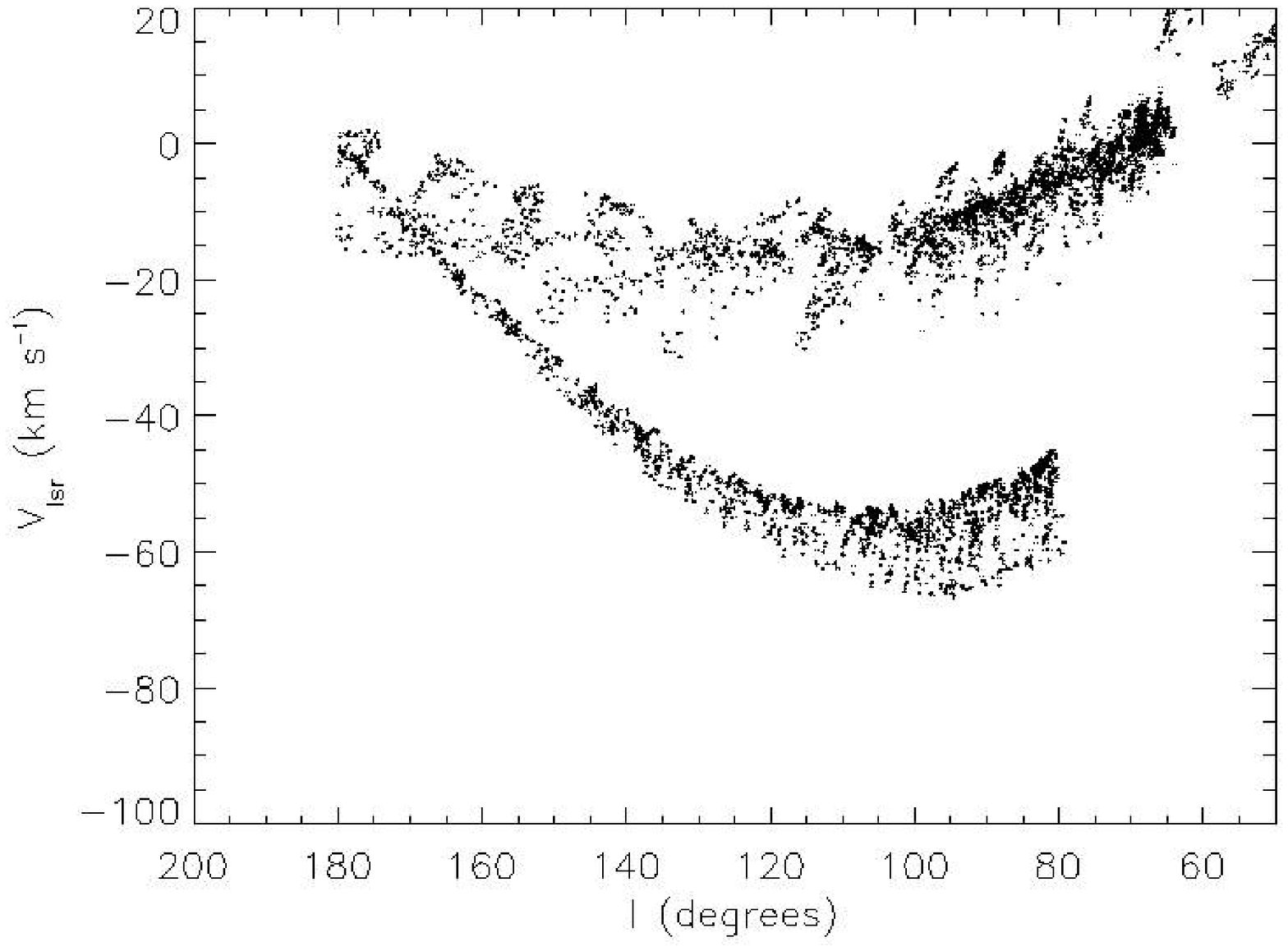,height=2.4in}}
\centerline{\psfig{file=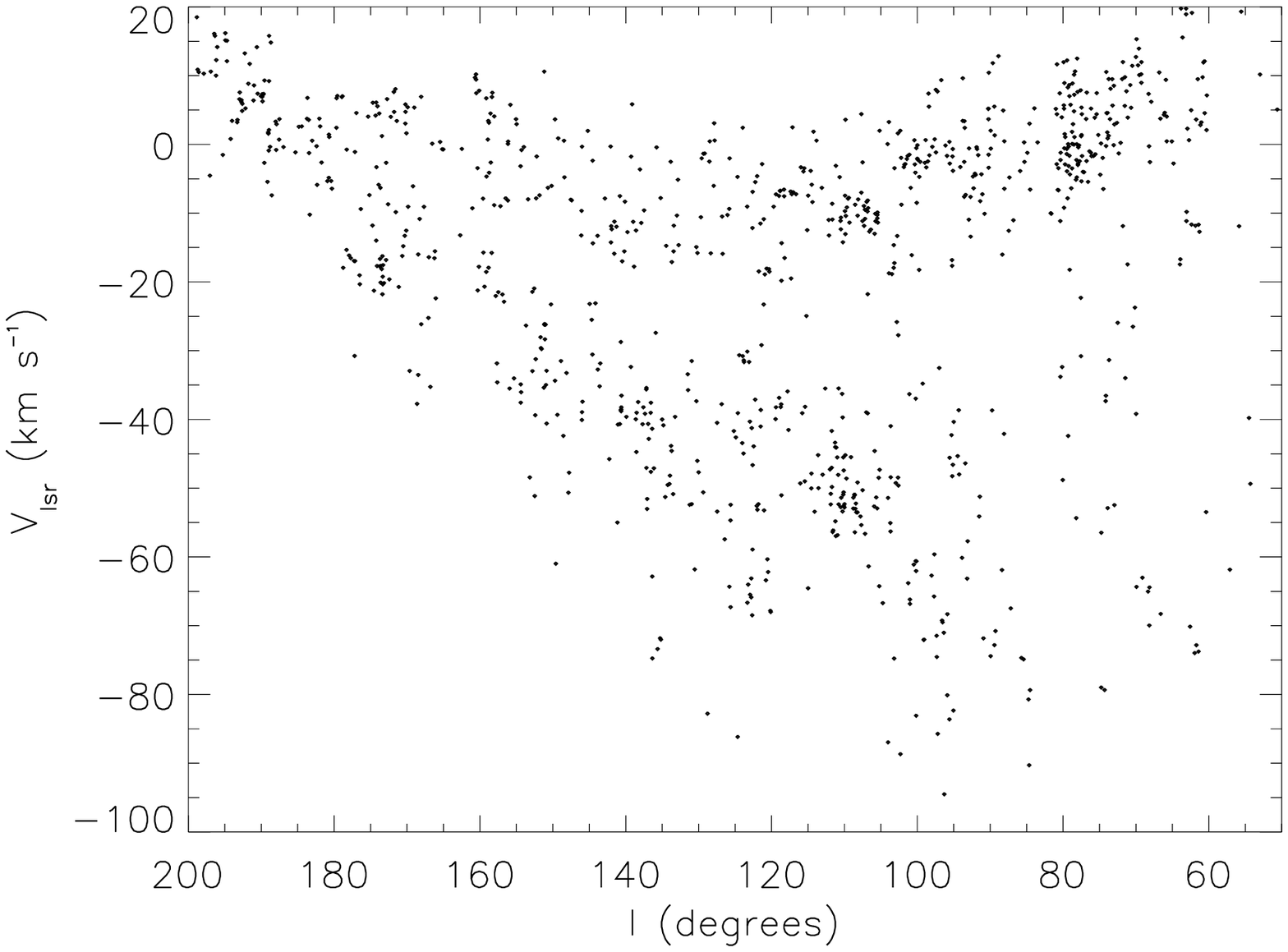,height=2.4in}}
\caption{V$_{lsr}$ plotted against galactic longitude for molecular clouds
from these simulations and the CO molecular line survey.
The top plot shows the simulated
data, with a LSR at coordinates (0,6.5)kpc. The lower plot uses data from
\citet{Yang2002}.}
\end{figure}

In Figures~14 \& 15, typical results from our calculations are compared with the FCRAO
CO survey and the CO molecular line survey (where the LSR is situated at
Cartesian coordinates (0,6.5)kpc). Our grid based method gave a total of 5044
objects within the sample criterion.  Below these
results corresponding data from the FCRAO CO survey are shown, where there are
7724 objects and the molecular line survey, where there are 1331 objects. 
The data from our model and the 2 surveys show clear bands, where 
the molecular clouds have
preferential radial velocities due to the spiral arm structure.
The molecular clouds in our model are more confined to the spiral arms than in
the observational data, particular for the outer arm. We have assumed a
corotation radius of 11~kpc, which may reduce the spiral arm 
shock at larger radii.    

Figure~16 displays the molecular clouds as viewed from outside the disk, at
coordinates (0,10.5)kpc. This is more comparable with Stark \& Lee (2005), who
observe clouds in the inner regions of the Milky Way. However their survey
incorporates 5 spiral arms, whereas only 2 are clear in Figure~16.
    
\begin{figure}
\centerline{\psfig{file=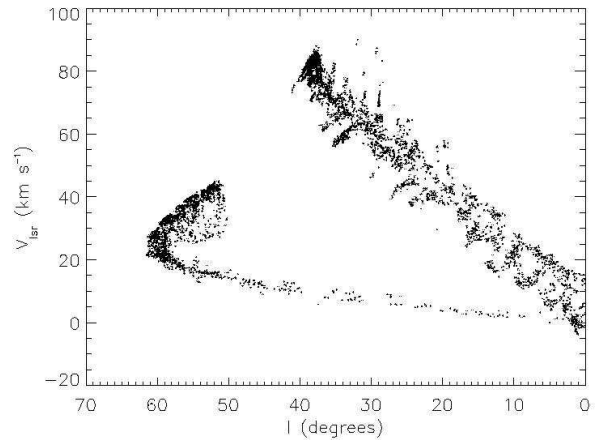,height=2.5in}}
\caption{V$_{rms}$ plotted against galactic longitude from simulated data. The
LSR was chosen at coordinates (0,10.5)kpc}
\end{figure}
\section{Conclusions}
The simulations in this paper demonstrate the formation of molecular clouds
in spiral galaxies, and specifically the role of spiral shocks. 
The compression
as the gas passes through a spiral shock produces high density regions
that are appropriate for the formation of molecular hydrogen. 
A necessary condition
for this to occur is that the gas is cold ($T\lesssim 100$~K). 
Warm gas ($T\ge 1000$~K) produces little molecular hydrogen.

The spiral shock also produces significant structure in the gas along the arms. Cold gas
is unstable to clumping in the shocks due to the semi-random encounters with lower and higher
angular momentum gas. These denser structures are predominantly molecular
and therefore identifiable as molecular clouds. An increased velocity dispersion
is apparent in the spiral arms and thus molecular clouds, of similar
magnitude to observations. We associate this increase in velocity dispersion 
with the generation of chaotic motions in the ISM as gas passes through 
clumpy shocks \citep{Bonnell2006}. We find only a small increase though in 
velocity dispersion in the vertical direction. Our simulations are not well 
resolved in the $z$ direction, and we would probably not expect to find
high velocities in this direction without stellar energy injection \citep{deAvillez2005}.

The total percentage of molecular hydrogen is typically 
$\sim 10\%$ for gas less
than 100~K and a total disk gas mass of $5~\times 10^8$~M$_{\odot}$,
(corresponding to a surface density of $\approx~2$~M$_{\odot}$~pc$^{-2}$), but
this figure roughly doubles with a
surface density of 20~M$_{\odot}$~pc$^{-2}$ or lower photodissociation rate. 
We note that these figures are potentially lower
limits on the amount of molecular hydrogen formed in these models, 
since we find that the total percentage of
molecular hydrogen increases with resolution. A caveat is that 
these results do not
include realistic heating and cooling, which are likely to have some effect.
 
Since molecular gas formation is determined by density, and the time spent in
dense regions, molecular gas is
largely confined to the spiral arms. Typically we find the ratio of arm to
interarm molecular hydrogen is 2 or 3 orders of magnitude. However a reduced
photodissociation rate allows the survival of molecular gas into the interarm
regions, and a ratio of arm to interarm molecular hydrogen less than 10. This
leads to the possibility of molecular gas passing from one arm to the other and
potentially determining subsequent molecular cloud formation. 

The distribution of molecular clouds in our simulations is consistent with 
observational surveys of molecular clouds in the Milky Way. The confinement of
molecular clouds to the spiral arms is more comparable to 
the outer Milky Way, where the ratio of
arm to interarm clouds is high. The resolution is insufficient to gain accurate
information about these clouds. Higher resolution
simulations will determine whether the properties of individual molecular clouds
(e.g. mass or internal velocity dispersion) can also be reproduced.  

\section*{Acknowledgements}
Computations included in this paper were performed using the UK Astrophysical
Fluids Facility (UKAFF). JEP thanks STScI for continuing support under the Visitor's
Program. We are grateful to Ron Allen, Bruce Elmegreen and Stephen
Lubow for enlightening discussions. We also wish to thank the referee for
helpful comments which improved the clarity of this paper.    
\bibliographystyle{mn2e}
\bibliography{Dobbs}

\bsp

\label{lastpage}

\end{document}